\documentclass[useAMS]{mn2e}
\usepackage{epsfig}
\newcommand{\beq}{\begin{eqnarray}}
\newcommand{\enq}{\end{eqnarray}}
\newcommand{\mateen}{   \left(\begin{array}{c}}
\newcommand{\mattwe}{   \left(\begin{array}{cc}}
\newcommand{\matdri}{   \left(\begin{array}{ccc}}
\newcommand{\matend}{   \end{array} \right)     }
\newcommand{\detdri}{   \left|\begin{array}{ccc}}
\newcommand{\detend}{   \end{array} \right|     }
\newcommand{\m}{\mathbf}

\newcommand{\eps}{\epsilon}

\newcommand{\Momega}{\mbox{\boldmath$\Omega$}}

\newcommand{\dotprod}{\mbox{\boldmath$\cdot$}}
\newcommand{\crossprod}{\mbox{\boldmath$\times$}}
\newcommand{\Evec}{\mbox{\boldmath$E$}}
\newcommand{\Bvec}{\mbox{\boldmath$B$}}
\newcommand{\bra}{\left[}
\newcommand{\ket}{\vphantom{\sqrt{0}} \right]}
\newcommand{\sn}{\ \! \mbox{sn}}
\newcommand{\cn}{\ \! \mbox{cn}}
\newcommand{\dn}{\ \! \mbox{dn}}

\newcommand{\hL}{\hat\m{L}}
\newcommand{\hO}{\hat\m{\Omega}}
\newcommand{\hb}{\hat\m{b}}

\newcommand{\Dsig}{\Delta\sigma}

\newcommand{\wc}{W_{\rm core}}
\newcommand{\icore}{I_{\rm core}}
\newcommand{\umin}{u_{\rm min}}
\newcommand{\he}{\hat\m{e}}
\newcommand{\sigi}{\sigma_i}
\newcommand{\cosi}{\cos\sigma_i}
\newcommand{\cosis}{\cos^2\sigi}
\newcommand{\sinis}{\sin^2\sigi}
\newcommand{\sini}{\sin\sigma_i}
\newcommand{\rhoi}{\rho_i}

\newcommand{\hx}{\hat\m{x}}
\newcommand{\deltar}{\delta_r}
\newcommand{\deltat}{\delta_t}

\title[Precession of the Isolated Neutron Star PSR B1828--11]
{Precession of the Isolated Neutron Star PSR B1828--11}
\author[T. Akg\"{u}n, B. Link and I. Wasserman]
{Taner Akg\"{u}n$^{1}$\thanks{E-mail: akgun@astro.cornell.edu;
link@physics.montana.edu; ira@astro.cornell.edu}, Bennett
Link$^{2,3}$ and Ira Wasserman$^{1}$\\
$^{1}$Center for Radiophysics and Space Research, Cornell University, Ithaca, NY 14853\\
$^{2}$Department of Physics, Montana State University, Bozeman, MT 59717\\
$^{3}$Department of Physics ``Enrico Fermi'', University of Pisa, Italy}
\begin{document}

%\date{Accepted 1066 December 7. Received 1988 December 14; in original form 1988 October 11}

\pagerange{\pageref{firstpage}--\pageref{lastpage}} \pubyear{2005}

\maketitle

\label{firstpage}

\begin{abstract}
Stairs, Lyne \& Shemar have found that arrival time residuals from PSR B1828--11 vary
periodically with a period $\approx 500$ days. This behavior can be accounted for by
precession of the radiopulsar, an interpretation that is reinforced by the detection of
variations in its pulse profile on the same timescale. Here, we model the period
residuals from PSR B1828--11 in terms of precession of a triaxial rigid body. We include
two contributions to the residuals: (i) the {\it geometric} effect, which arises because
the times at which the pulsar emission beam points toward the observer varies with
precession phase; (ii) the {\it spindown} contribution, which arises from any dependence
of the spindown torque acting on the pulsar on the angle between its spin ($\hO$) and
magnetic ($\hb$) axes. We use the data to probe numerous properties of the pulsar, most
notably its shape, and the dependence of its spindown torque on $\hO\dotprod\hb$, for
which we assume a sum of a spin-aligned component (with a weight $1-a$) and a dipolar
component perpendicular to the magnetic beam axis (weight $a$), rather than the vacuum
dipole torque ($a=1$). We find that a variety of shapes are consistent with the
residuals, with a slight statistical preference for a prolate star. Moreover, a range of
torque possibilities fit the data equally well, with no strong preference for the vacuum
model. In the case of a prolate star we find evidence for an angle-dependent spindown
torque. Our results show that the combination of geometrical and spin-down effects
associated with precession can account for the principal features of PSR B1828--11's
timing behavior, without fine tuning of the parameters.
\end{abstract}


\section{Introduction}

Pulse arrival times of neutron stars can be found very accurately, which allows for the
determination of the spin period and period derivative to very high precision. Normally,
the time of arrival \emph{residuals} which are calculated by subtracting the period and
the period derivative (and in some cases the period second derivative) are mostly white
noise. However, residuals from a small number of rotating neutron stars are found to
exhibit long term cyclical, but non-oscillatory, variations with characteristic
timescales of order months to years (Cordes 1993). The variability may be temporary (e.g.
the Vela pulsar during its Christmas glitch [McCulloch et al. 1990]) or persistent (e.g.
the accreting neutron star Her X-1 [Tannanbaum et al. 1972], the Crab pulsar [Lyne,
Pritchard and Smith 1988], and the pulsars  PSR 1642-03 [Blaskiewicz 1992], PSR B0959-54
[D'Alessandro and McCulloch 1997] and PSR B1828--11 [Stairs, Lyne and Shemar 2000]). The
long timescales that characterize the observed variations would arise naturally from
precession\footnote{Throughout this paper, we call this phenomenon precession, as has
become common in the literature, although purists might prefer the term nutation.}, when
the principal axes of a body (defined through the moments of inertia, which we will take
as $I_1 \leq I_2 \leq I_3$) revolve periodically around the angular momentum, as viewed
in an inertial frame. An ellipticity $\epsilon = (I_3 - I_1)/I_1 \ll 1$ would be expected
to produce variations in the timing residuals of an axisymmetric body with a period
$P_p=P_\star/\epsilon\approx 3.2\,P_\star({\rm sec}) (10^8\epsilon)^{-1}$ years, where
$P_\star$ is the rotation period. The arrival time variations characteristic of
precession would be strictly periodic, but not sinusoidal for a triaxial rotator.

\begin{figure}
        \centerline{\includegraphics[scale=0.415]{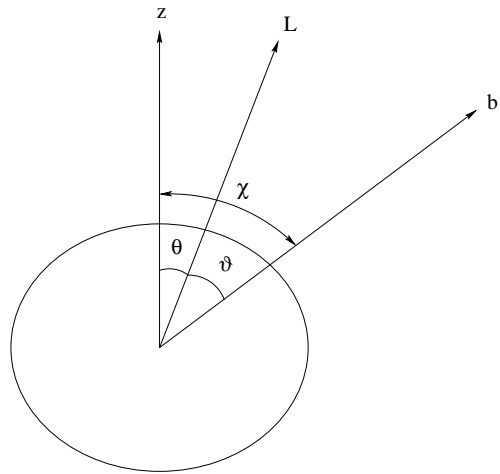}}
        \caption{Definition of various angles: the wobble angle ($\theta$) is the angle
        between the angular momentum ($\hL$) and the body $z$ axis (which is chosen as
        the principal axis corresponding to the largest moment of inertia, $I_3$);
        the beam swing angle
        ($\vartheta$) is the angle between the angular momentum and the magnetic axis
        ($\hb$); $\chi$ is the polar angle of the magnetic axis in the body frame.
        Note that the angles may not be coplanar. For an axisymmetric body the wobble
        angle remains constant; for a triaxial body it varies with time (see Appendix B).}
        \label{angles}
        \end{figure}

There are two physical causes for time of arrival residuals ($\Delta t$) in a
precessing neutron star (Cordes 1993). One is directly geometrical: as the
rotating star precesses, the symmetry axis of its radiation beam crosses the
plane defined by the angular momentum of the star and the direction to the
observer at times that vary periodically over the precession cycle. The magnitude
of the variability is set by the amplitude of the precession, which is roughly
the \emph{wobble angle}, $\theta$ (defined as the angle between the angular
momentum and the principal axis corresponding to the largest moment of inertia,
Fig. \ref{angles}). Typically $\theta\ll 1$, and the amplitude of the time of
arrival residuals is $\Delta t_{\rm geo} \sim\theta P_\star$. In addition, the
dependence of the spindown torque acting on the pulsar on the angle between its
spin and magnetic axes produces a timing residual that can be comparable to and
even exceed $\Delta t_{\rm geo}$. If we assume that the spindown torque is
proportional to $\hO - a(\hO\cdot\hb)\hb$, where $\Momega=\Omega\hO$ is the
angular velocity, $\hb$ is the magnetic axis, and the dimensionless parameter $a$
is a measure of the angular dependence ($a\equiv 1$ for a spinning magnetic
dipole radiating into vacuum) then the spindown rate varies over the precession
cycle as well, producing a timing residual $\Delta t_{\rm sd}\sim a\theta
P_p^2/t_{\rm sd} \sim (aP_p^2/P_\star t_{\rm sd})\Delta t_{\rm geo}$, where
$t_{\rm sd}$ is the spindown timescale for the pulsar; the dimensionless
parameter $\Gamma_{\rm sd}\equiv P_p^2/P_\star t_{\rm sd} \approx 3.2\,P_p^2({\rm
years})/P_\star({\rm sec}) (t_{\rm sd}/10^7~{\rm years})$ may be large.
Associated with these arrival time residuals are period residuals $(\Delta
P/P_\star)_{\rm geo}\sim\theta P_\star /P_p\approx 3.2\times 10^{-8}\theta
P_\star({\rm sec})/P_p({\rm years})$ and $(\Delta P/P_\star)_{\rm sd}\sim
a\Gamma_{\rm sd}(\Delta P/P_\star)_{\rm geo}$.

\begin{figure}
        \centerline{\includegraphics[scale=0.355]{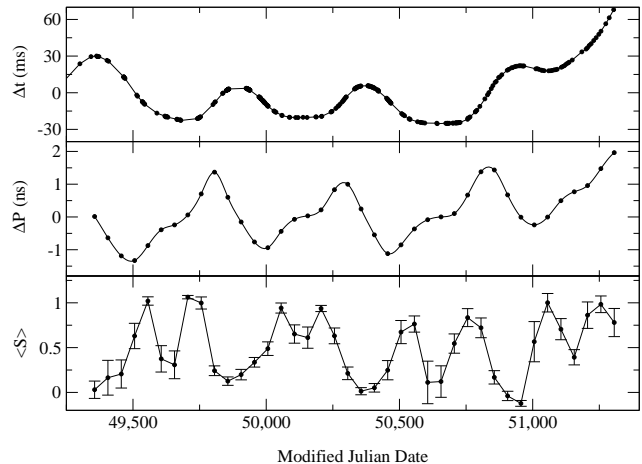}}
        \caption{Time of arrival residuals, period residuals, and shape parameter
        for PSR B1828--11 (courtesy of I. H. Stairs, A. G. Lyne and S. L. Shemar).
        The shape parameter is defined in terms of the weight of the narrow ($A_N$)
        and wide ($A_W$) standard pulse profiles that are present at every epoch,
        $S = A_N / (A_N + A_W)$, so that $S \simeq 1$ for narrow pulses, and
        $S \simeq 0$ for wide ones (Stairs, Lyne \& Shemar 2000; Stairs et. al. 2003).
        The solid line uses a cubic spline to connect the points for $\Delta P$.}
        \label{data}
        \end{figure}

The best candidate to date for truly periodic long term variations in arrival times is
PSR B1828--11 (Stairs, Lyne \& Shemar 2000; Stairs et. al. 2003). Fourier analysis of
these variations reveals harmonically related periodicities at approximately $1000$,
$500$ and $250$ days (Stairs, Lyne \& Shemar 2000; Fig. \ref{data}), with the latter two
somewhat more pronounced than the first. The length of the timescale of these variations
implies that they are probably not of magnetospheric origin, since the natural timescale
in the magnetosphere is of the order of the spin period, which in this case is $P_\star =
0.405$ sec. Even the $\Evec\crossprod\Bvec$ drift of subpulses (e.g. Ruderman \&
Sutherland 1975) does not exceed $\sim 10 P_\star$. (However, Ruderman [2001] has
suggested the possibility of drifts with periods of the order of a year.) As of the time
of writing, there are no quantitative models for the data based solely on magnetospheric
effects, but there are successful models based on precession (e.g. Jones \& Andersson
2001, Link \& Epstein 2001, and Wasserman 2003). Link \& Epstein (2001) previously
modelled the timing residuals from this pulsar in terms of precession of an axisymmetric,
oblate rotating rigid body slowing down according to the vacuum magnetic dipole radiation
formula. They found that the observations could be accounted for in this model provided
that the underlying pulsar is nearly an orthogonal rotator (magnetic obliquity
$\chi\approx 89^\circ$ to the body's symmetry axis) and nearly aligned angular momentum
(wobble angle $\theta\approx 3^\circ$ between angular velocity and symmetry axis). These
are in accordance with the conclusions reached by Jones \& Andersson (2001). Although the
precession amplitude is small, it may suffice to unpin superfluid vortex lines (Link \&
Cutler 2002), thus avoiding a potential impediment to precession: pinning was shown to
shorten the precession period to about 100 spin periods, and precession itself is
dissipated over a timescale of 100-10,000 precession periods (Shaham 1977, 1986;
Sedrakian, Wasserman \& Cordes 1999). Wasserman (2003) argued that the data could also be
accounted for if the underlying neutron star has either a type II superconductor or a
strong toroidal magnetic field in its core. In these models, the angle between the
angular velocity and symmetry axis of the star could be larger than found by Link \&
Epstein (2001), and the star does not have to be a nearly orthogonal rotator; spindown
variations were found to dominate the timing residuals in this case as well. Link (2003)
showed that the standard picture of the core of type II superconducting protons
coexisting with superfluid neutrons is inconsistent with long-period precession; pinning
of the neutron vortices to the proton flux tubes makes the precession frequency
comparable to the rotation frequency of the star, a factor of $10^8$ too fast.  Possible
implications include a normal core (both neutrons and protons), superfluid neutrons and
normal protons, normal neutrons and superfluid protons (type I or type II), or superfluid
neutrons and type I protons. Sedrakian (2005) studied the last possibility. He calculated
the drag on neutron vortices moving in a type I superconducting core, and found that the
drag is sufficiently small (that is, the vortices are sufficiently mobile with respect to
the protons) that long-period precession is indeed possible in this scenario.
Irrespective of the details, magnetic stresses in excess of the relatively weak ones that
would arise from the pulsar's apparent dipole field strength, together with crustal
stresses, would render the neutron star effectively triaxial in shape (Cutler 2002;
Wasserman 2003; Cutler, Ushomirsky \& Link 2003).

Link \& Epstein (2001) and Wasserman (2003) gave two alternative models that interpret
the timing of PSR B1828--11 as precession. These models fit the data well, thus providing
strong evidence that the observed timing variations do indeed represent precession. These
two models, however, are special cases. The purpose of this paper is to do a thorough
search of the parameter space to see what we can learn about the properties of the
spindown torque and the stellar figure. To this end, we analyze the {\it period}
residuals from this pulsar in terms of a simple model in which the rotating neutron star
is assumed to be a {\it triaxial rigid body}. Obviously, precise axisymmetry is a special
case, and we do not expect it to hold generally, particularly if the crust of the star is
not in a relaxed state (e.g. Cutler, Ushomirsky \& Link 2003), or has substantial
internal magnetic stresses that may not be axisymmetric to begin with. Thus, one of our
goals is to see what the data from PSR B1828--11 reveal about the shape of the neutron
star crust.

In this paper, we model a precessing neutron star as a single (rigid) body, rotating
uniformly. Realistic neutron star modelling should take account of at least two different
components -- its solid crust, and (super)fluid core. Bondi \& Gold (1955) considered the
precession of a body consisting of a solid crust coupled frictionally to a fluid core.
Their work showed that the long-term precession of the composite system depends on the
timescale $t_{cc}$ on which the crust and core couple to one another. If $\Omega
t_{cc}\ll 1$ then the crust and core are very tightly coupled to one another on
timescales smaller than a rotation period, and the moment of inertia tensor relevant to
precession is that of the entire system, crust plus core. In this case, precession damps
out slowly, on a timescale $\sim (\Omega t_{cc})^{-1}$ {\it precession} periods. If
$\Omega t_{cc}\gg 1$ the crust and core only couple on timescales long compared to a
rotation period, and the moment of inertia tensor relevant to precession is that of the
crust alone. In this case, precession also damps slowly, with a characteristic decay
timescale $\sim\Omega t_{cc}$ precession periods. Estimates of the crust-core coupling
timescale vary but the consensus is that the coupling is weak, with $\Omega t_{cc}\sim
10^2-10^4\gg 1$ (e.g. Alpar, Langer \& Sauls 1984, Alpar \& Sauls 1988, Sedrakian \&
Sedrakian 1995), so the precession dynamics are governed by the moment of inertia tensor
of the crust alone. However, it also turns out that as long as the crust and core couple
on a timescale short compared with a precession period, but still long compared to the
spin period, the relevant moment of inertia for {\em all spindown effects, including
those that vary periodically over a precession period, is the total stellar moment of
inertia} (Akgun, Link \& Wasserman 2005, in preparation); this is the appropriate regime
as long as $\Omega t_{cc}\ll 10^8$, which appears to be the case. Thus, our one component
model for precession is justified, apart from slow decay of the precession, which we
neglect.

Another issue is that the neutron star crust is not perfectly rigid, but has a finite
shear modulus. For a biaxial precessing star, the crust must be strained in order for the
star to precess with a period of order a year (Cutler, Ushomirsky \& Link 2003). In
addition, the strain field will vary with time as the star precesses, making the star's
moment of inertia tensor time-dependent in the body frame. For simplicity, we neglect
these effects, and assume that the rotation of an imperfectly rigid, triaxial star is
well-described by the Euler equations for a rigid body, but with a moment of inertia
tensor that is rescaled to account for the finite shear modulus.

Since $P_\star=0.405$ seconds and $t_{\rm sd}\approx 10^5$ years for PSR B1828--11,
$\Gamma_{sd}\sim 10^3$, and the spindown contribution to the precession-induced timing
residuals is particularly important. As a result, we may also hope to use these data to
probe the value of $a$, that is, to probe the angular dependence of the spindown torque.
While it is common to assume that $a\equiv 1$ for rough analysis, on theoretical grounds
we should not expect this to be true, for even an aligned rotator surrounded by a
magnetosphere radiates energy, a process whose source is ultimately the rotational energy
of the star, resulting in spindown at a rate presumably not much different from its
luminosity divided by its rotational frequency. One of our chief findings is evidence
that the external torque that spins down a pulsar does indeed possess at least some
angular dependence.

Our analysis uses the same segment of the data\footnote{We thank I. H. Stairs, A. G. Lyne
and S. L. Shemar for generously sharing their timing residual data with us.} that was the
basis of Link \& Epstein (2001); this facilitates direct comparisons between the results
of the two studies. We focus on the period residuals because we can derive an analytic
formula for them in terms of elliptic functions. (An analogous formula for an oblate
axisymmetric rotator has been derived previously by Bisnovatyi-Kogan, Mersov \& Sheffer
1990, and Bisnovatyi-Kogan \& Kahabka 1993, and has been applied to the 35-day cycle of
Her X-1.) Because the underlying triaxial model involves numerous parameters, using an
analytic formula speeds up the computation considerably, which is a distinct advantage.
By contrast, direct analysis of the timing residuals would require numerical integration
of the model equations, a distinct disadvantage. Thus, for computational convenience, we
analyze the period residuals rather than the arrival time residuals. However, a
straightforward analysis of the period residuals using their tabulated uncertainties
yielded very large values of $\chi^2$ ($\sim {\rm a~few~} \times 10^{3}$) under the
assumption that the residuals are due solely to precession. This indicated to us that
there is extra noise in the period residuals, either because their estimated
uncertainties are too small (which we regard as unlikely to account for all the noise) or
because there is a physical source of period noise that smears out the smooth
contribution from precession systematically. In order to account for this extra noise
simply, we multiplied the tabulated uncertainties by a (single) factor $F$, and then
marginalized over $F$ to obtain posterior distributions of the (more interesting)
parameters of the precession model. This method -- Student's t-test -- represents a
computational realization of ``chi-by-eye'' for data whose uncertainties may only be
known incompletely. Details are given in Appendix D.

Section 2 contains basic features of our model; further details may be found in
Appendices A, B and C. Section 3 contains results and implications of our analysis;
statistical details, including the ``chi by eye'' method mentioned above, are found in
Appendix D. Section 4 is a short digression on the pulse shape of PSR B1828--11, which is
seen to vary systematically with precession phase (Stairs et al. 2000). Although we do
not use this information in our statistical analysis, precession samples different
regions in a pulsar's radio-emitting region, and offers the possibility of mapping out
its shape (as has been done for PSR 1913+16, which exhibits geodetic precession, by
Weisberg, Romani \& Taylor [1989], and previously by Link \& Epstein [2001] for PSR
B1828--11).

\section{Models}

Here we review the models that we use briefly, highlighting some of the more important
parameters. The derivations for the period residuals are lengthy and are left for the
Appendices; we present the geometric model in Appendix A, and the spindown model is
derived in Appendix C. In what is next, we will follow the notation used in these
Appendices. The parameters of our models are listed in Table \ref{table1}.

        \begin{table}
        \caption{Definitions of important parameters.}
        \label{table1}
        \begin{tabular}{ll}
        \hline
        parameter &
        \begin{tabular}{l}
        physical meaning
        \end{tabular} \\ \hline
        \vspace{0.17cm} $\chi$, $\phi$ &
        \begin{tabular}{l}
        polar and azimuthal angles of the \\ magnetic axis in the body frame
        \end{tabular} \\
        \vspace{0.17cm} $e^2$ &
        \begin{tabular}{l}
        measures the degree of triaxiality
        \end{tabular} \\
        \vspace{0.17cm} $\lambda$ &
        \begin{tabular}{l}
        determines the components of the \\ angular momentum and is \\
        related to the wobble angle
        \end{tabular} \\
        $a$ &
        \begin{tabular}{l}
        determines the strength of the angular \\ part of the spindown torque
        \end{tabular} \\ \hline
        \end{tabular}
        \end{table}

\subsection{Geometric Model}

What we refer to as the \emph{geometric} model is the effect of triaxiality alone (i.e.
torque-free precession). In this case, Euler's equation for the angular momentum can be
solved analytically in terms of Jacobian elliptic functions (Landau \& Lifshitz, 1976).
As we show in Appendix A, the period residuals are then found to be of the form $\Delta
P_{ge} / P_\star \approx \varpi_p f_n$ where $P_\star$ is the rotation period at a
fiducial epoch, $\varpi_p$ is a dimensionless quantity of order $P_\star/P_p \sim
\epsilon$, and $f_n$ is a complicated combination of the elliptic functions. Because of
the inherent form of $f_n$ the amplitude of the residuals is not trivial to predict in
general, and they can exhibit very rich behavior.

We denote the principal moments of inertia by $I_i$; the associated axes serve as the
basis for the rotating (body) frame. We then define the following parameters: $\epsilon =
(I_3-I_1)/I_1$ which measures the deviation from sphericity; $e^2 =
[I_3(I_2-I_1)]/[I_1(I_3-I_2)]$ which measures the degree of triaxiality; $k^2$ which is
the parameter of the Jacobian elliptic functions, and depends on the angular momentum and
the moments of inertia; and $\lambda$ which determines the components of the angular
momentum (and would be simply $\lambda = L_1/L_3$ for an axisymmetric star, but is
slightly different in the more general case, see Eq. (\ref{lambda})). The last three are
not independent: $k = e \lambda$. Note that $k^2$ does not depend on $\epsilon$, but only
on $e^2$ and $\lambda$.

        \begin{figure}
        \centerline{\includegraphics[scale=0.37]{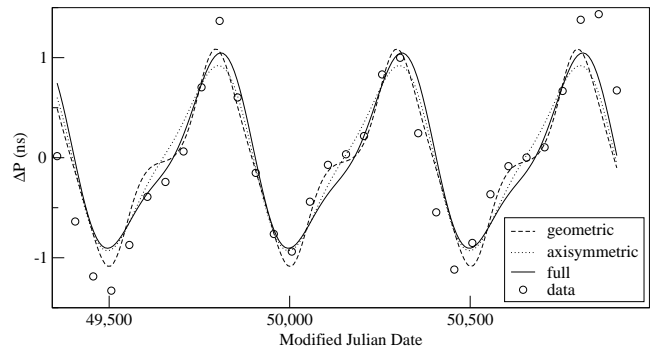}}
        \caption{Sample models for the period residuals. The data points
        are shown for comparison. The solid line is for the full model (both effects
        included), the dashed line is for the geometric model alone, and
        the dotted line is for the axisymmetric model. The
        parameters for all models shown here are quoted in Table \ref{table}.}
        \label{plots}
        \end{figure}

        \begin{table}
        \caption{List of parameters for the models shown in Fig. \ref{plots}.}
        \label{table}\hspace{0.1cm}
        \begin{tabular}{cccc}
        \hline
        parameter & geometric & axisymmetric & full \\ \hline
        $\chi$ & $74.0^\circ$ & $88.6^\circ$ & $71.8^\circ$ \\
        $\phi$ & $12.0^\circ$ & $0^\circ$ & $0^\circ$ \\
        $e^2$ & 3912 & 0 & 2135 \\
        $\lambda$ & 0.012 & 0.0437 & 0.00325 \\
        $a$ & 0 & 1 & 0.983 \\ \hline
        $\Delta\vartheta$ & $18.5^\circ$ & $5^\circ$ & $0.44^\circ$ \\ \hline
        \end{tabular}
        \end{table}

Note that for $e^2 = 0$ the body is \emph{oblate} and axisymmetric, and $\lambda = 0$
when there is no precession. In both cases $k = 0$, and the Jacobian elliptic functions
reduce to the regular trigonometric functions. At $k = 1$ they become hyperbolic
functions, and the angular momentum exponentially aligns with the principal axis
corresponding to the intermediate moment of inertia, $I_2$. Between these two extremes,
the precession of the angular momentum takes place along the intersection of the sphere
defined by the conservation of angular momentum ($L^2 = L_iL_i$), and the ellipsoid
defined by the conservation of energy ($E = L^2_i/2I_i$). The resulting shape of the
trajectories is the Binet ellipsoid (Landau \& Lifshitz, 1976). In the limit $e^2
\rightarrow \infty$ the body becomes \emph{prolate} axisymmetric.

\subsection{Spindown Model}

The rotation of an isolated neutron star is not torque-free, but slows down, resulting in
a gradual increase in the rotation period. It is thought that because of the rotating
magnetic field, angular momentum is lost to radiation. In the simple model of a rotating
dipole in a vacuum, the pulses are emitted at the poles of the magnetic field and the
torque has the form $\m{N} \sim - N_o[\hO - (\hO\cdot\hb)\hb]$, where $\hO$ is the
instantaneous rotation axis, and $\hb$ is the pulse (and magnetic) axis.

This is a very crude model. The magnetic field may have non-dipolar components of
considerable amplitude, which we will not consider here. The pulsar is also not in a
perfect vacuum, but is surrounded by a plasma-filled magnetosphere (Goldreich \& Julian,
1969). The vacuum torque vanishes when the angular velocity and the magnetic axis are
aligned, while the presence of a magnetosphere would require a loss of angular momentum,
no matter what the orientation is. Therefore, the vacuum dipole torque should give only
an incomplete description at best. We adopt a general spindown torque of the form
$\m{N}_{sd} = - N_o [\hO - a (\hO\cdot\hb)\hb]$, where we introduce an additional
parameter $a$. Loss of angular momentum mandates that $a \leq 1$. It should also be
positive, or we would have an angle dependent torque that is opposite in sign to the
dipole contribution. The vacuum case is retrieved by setting $a=1$. The case of $a = 0$
corresponds to an external torque with no angular dependence, which would not produce
periodic time of arrival residuals.

The torque-modified Euler's equation can be solved approximately for small $\epsilon$,
and a second contribution to the period residuals arises due to the torque, $\Delta
P_{sd} / P_\star \approx \tilde{\Gamma}_{sd}\Delta\tilde\ell$. We will refer to this as
the \emph{spindown} model; and the sum of both geometric and spindown contributions will
be referred to as the \emph{full} model. Here, $\tilde\Gamma_{sd} = I_3 N_o / \varpi_p
L^2 \sim P_p / t_{sd} \sim 10^{-5}$ and is determined by the spindown properties of the
neutron star (in particular, the period derivative, $\dot P$ or the characteristic age,
$t_{sd}$); $\Delta\tilde\ell$ is another complicated function of Jacobian elliptic
functions and Legendre integrals. For an axisymmetric star, $\Delta\tilde\ell = a_1 \sin
\omega_p t + a_2 \sin 2\omega_p t$, where $\omega_p$ is now the precession frequency, and
$a_i$ are some coefficients; the axisymmetric model thus has two harmonically related
components. We take the two peaks in the spectrum of the period residuals of PSR
B1828--11 with periods of $\sim 500$ and $\sim 250$ days to be the most significant. From
the form of the axisymmetric model, we thus conclude that the precession period must be
$\sim 500$ days. It can also be shown that for an axisymmetric star, in the region of
interest, the geometric contribution is quite negligible compared to the spindown
contribution for $a \sim 1$ (see Appendix C).

If the 1000--day period represented the precession period, we would expect to see
variations of the pulsar beam width at the same period. While such changes were reported
by Stairs, Lyne \& Shemar (2000), subsequent more detailed analysis has not shown a
1000--day period in the beam width data (Parry et al. 2005). We thus assume that the
precession period is $\sim 500$ days, and attribute the 1000--day component in the timing
data to something unrelated to precession, such as timing noise. We note that our model
cannot provide satisfactory fits to the data if the precession period is $\sim 1000$
days.

\subsection{Constraints and Statistical Analysis}
The orientation of the angular momentum, $\hL$ is fixed in the inertial frame. This is
still true even in the presence of the spindown torque, if $\epsilon$ is sufficiently
small (see Appendix C). Then the requirement that the pulse beam, which we assume to be
centered along the magnetic axis, $\hb$ never precesses entirely out of our line of sight
means that the angle between $\hL$ and $\hb$ should not vary by more than the angular
width of the pulse itself. We will refer to the angle between $\hL$ and $\hb$ as the
\emph{beam swing angle} and denote it by $\vartheta$. If the angular radius of the pulse
is $\rho$, then the above constraint can be expressed as $\Delta\vartheta =
\vartheta_{max} - \vartheta_{min} \leq 2\rho$; in general, we will require the beam swing
variation to be less than some value $\Delta\vartheta_{max}$. We also will define the
\emph{wobble angle}, $\theta$ as the angle between the angular momentum and the body $z$
axis (see Fig. \ref{angles} and Appendix B).

The duty cycle allows us to estimate the angular extent of the pulse, and for PSR
B1828--11 this varies between $5^\circ$ to $7^\circ$. For a circular pulse, this implies
that the beam swing angle cannot be varying by more than a few degrees. Larger variations
would require a more elongated pulse shape. Yet, at this time, there is not enough
evidence to elaborate more on this. In particular, polarization data might be quite
useful to determine the extent of the pulse. Stairs et al. (2000) also report periodic
variations in the average pulse shape. We offer a possible explanation in a following
section.

Another restriction may be that PSR B1828--11 does not have an interpulse. That can
further restrict the relative orientation of the angular momentum and the magnetic axis.
However, due to uncertainties in the structure of the magnetic field, it is not clear
that an interpulse will necessarily appear. Therefore, we do not impose this restriction.
The observer's location is an additional parameter, and can be independently fixed.

We apply the two models - geometric $(a = 0)$ and full $(a \neq 0)$ - under the given
constraints to PSR B1828--11, using a Bayesian approach to obtain probability
distribution functions (pdfs) for individual parameters. We assume specific priors in the
full multi-dimensional parameter space (to be discussed next), but the effective priors
exhibited in the projected (marginalized) 1-D posterior pdfs shown in the figures below
are integrals over the constraints. Because there appears to be systematic noise in the
period residuals larger than their tabulated uncertainties, we scale the latter and then
marginalize over the scaling factor as detailed in Appendix D. Once the likelihood is
determined, the individual pdfs are obtained through integration over the remaining
parameters and normalization.

\begin{figure}
        \centerline{\includegraphics[scale=0.34]{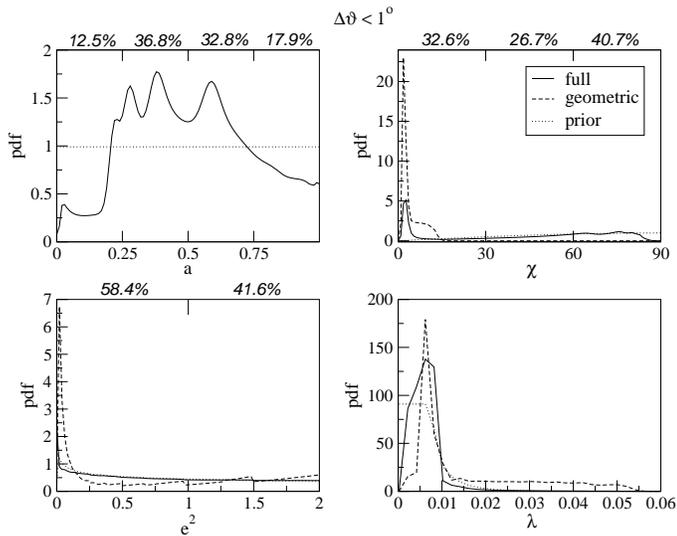}}
        \caption{Pdfs for beam swing angle variation less than $1^\circ$.
        In Figs. \ref{pdf1A}-\ref{pdf5A} the prior for $\chi$ is flat over
        $d(\cos\chi)$; all other priors are flat. The cutoff for $e^2$ is at 2, and
        $\lambda$ is confined to be below 0.2. $\chi$ is in degrees; $a$, $e^2$
        and $\lambda$ are dimensionless. The percentages listed above the plots are the
        probabilities for the full model enclosed in the corresponding ranges.}
        \label{pdf1A}
        \end{figure}

\begin{figure}
        \centerline{\includegraphics[scale=0.34]{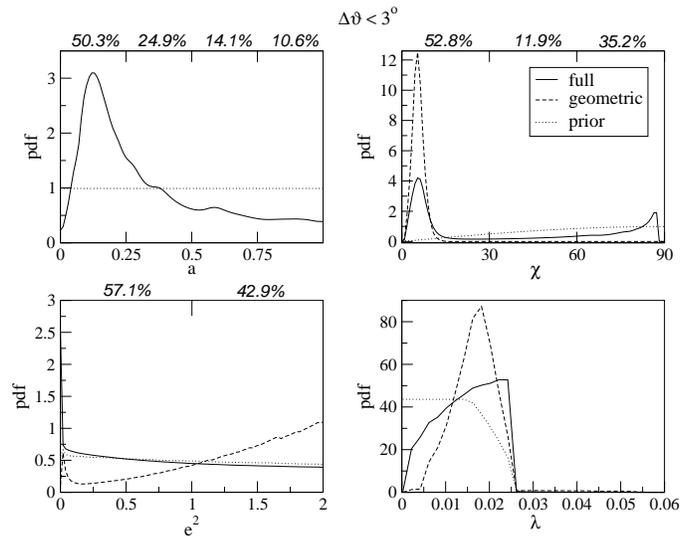}}
        \caption{Pdfs for beam swing angle variation less than $3^\circ$, for the same
        set of priors as in Fig. \ref{pdf1A}.}
        \label{pdf3A}
        \end{figure}

\begin{figure}
        \centerline{\includegraphics[scale=0.34]{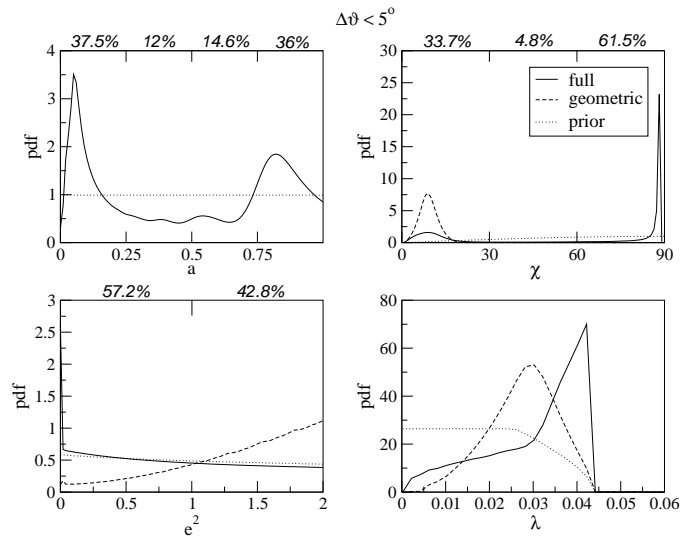}}
        \caption{Pdfs for beam swing angle variation less than $5^\circ$, for the same
        set of priors as in Fig. \ref{pdf1A}.}
        \label{pdf5A}
        \end{figure}

Let $\!\{p_k\!\}$ denote the set of the $n$ parameters. Then the likelihood, ${\cal
L}(\{p_k\!\})$ and the volume of integration, ${\cal V}(\{p_k\!\})$ are functions of this
set. The latter also depends on the beam swing angle constraint, which itself is a
function of a subset of the parameters. The priors, $g_i(p_i)$ are functions only of the
single parameter they refer to. Then the projected 1-D posterior pdf for the $i$-th
parameter can be expressed as an integral of the likelihood over the remaining
parameters, over the volume defined by the constraints,
        \beq
        f_i (\{p_k\!\}) = \int_{{\cal V}(\{p_k\!\})}
        g_i(p_i) {\cal L}(\{p_k\!\}) \prod_{j \neq i}^{n} g_j(p_j) \, dp_j \, .
        \enq
Similarly, the projected 1-D prior can be expressed as,
        \beq
        h_i (\{p_k\!\}) = \int_{{\cal V}(\{p_k\!\})}
        g_i(p_i) \prod_{j \neq i}^{n} g_j(p_j) \, dp_j \, .
        \enq
It is these two quantities ($f_i$ and $h_i$) that are plotted in Figs.
\ref{pdf1A}-\ref{pdf5B}. Note that, if the volume of integration had not depended on the
constraints, then we would simply have $h_i = g_i$.

Within the context of our precession model we can use the pdf to compare how well
different sets of model parameters fit the data. However, we cannot assess the extent to
which the data demand explanation in terms of precession, as opposed to some other,
completely different physical model. Any model for the data will lead to a pdf with local
maxima at certain values of the parameters of the model, and we can assess the relative
significance of these peaks to quantify the extent to which the model parameters are
determined by the data. Whether or not another model that is just as well-motivated
physically as our precession model can fit the data better is outside the scope of our
analysis. Given a competitor model - of which we are unaware - Bayesian methods could be
used for making model comparisons.

\section{Results and Discussion}

The physical parameters that determine the form of the residuals are the two angles that
specify the orientation of the magnetic axis in the body frame (the polar angle, $\chi$
and the azimuthal angle, $\varphi$; see Fig. \ref{pulsaxis}); any two of $e^2$, $\lambda$
and $k^2$; and $a$. There is also a $\tau_o$ (measured in precession cycles) that
determines the initial phase. Thus, the total number of parameters is six. $\chi$ varies
between 0 and $\pi/2$, and its prior is taken to be flat over $\cos\chi$; and $\phi$
varies between 0 and $2\pi$, and has a flat prior. In other words, we assume that
orientations of the magnetic axis are equally likely over all solid angles. Priors for
$\tau_o$ and $a$ are flat between 0 and 1. On the other hand, $e^2$ and $\lambda$ can
have any positive values, as long as the beam swing angle is constrained and $k^2 < 1$;
therefore, we have to introduce cutoffs in their priors. $\lambda$ is related to the
wobble angle, and due to the beam swing angle constraint it cannot be too large; we take
$\lambda \leq 0.2$ with a flat prior.

\begin{figure}
        \centerline{\includegraphics[scale=0.34]{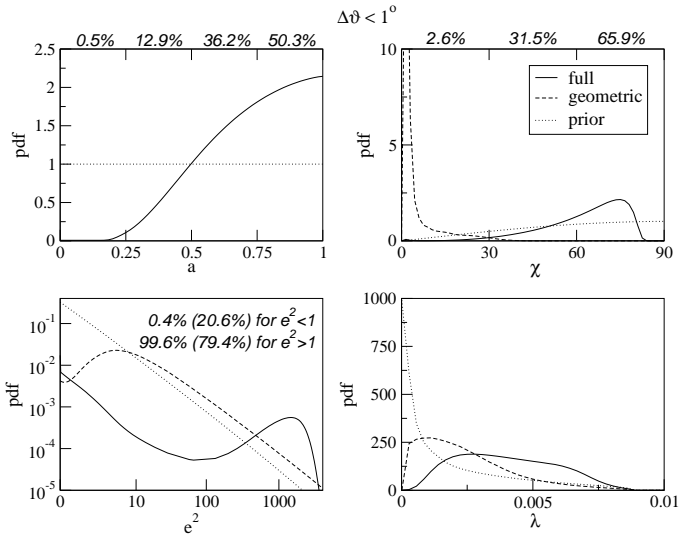}}
        \caption{Pdfs for beam swing angle variation less than $1^\circ$.
        In Figs. \ref{pdf1B}-\ref{pdf5B} the prior for $\chi$ is flat
        over $\cos\chi$; the prior for $e^2$ is flat over $\ln(1+e^2)$;
        all other priors are flat. The cutoff for $e^2$ is at 4000, and
        $\lambda$ is confined to be below 0.2. $\chi$ is in degrees; $a$, $e^2$
        and $\lambda$ are dimensionless. $e^2$ is plotted on a log-log scale to
        reveal more detail. The percentages listed above the plots are the
        probabilities for the full model enclosed in the corresponding ranges. The prior
        probabilities are given for $e^2$ in parentheses for comparison with the posterior
        probabilities.}
        \label{pdf1B}
        \end{figure}

The situation is slightly more complicated for $e^2$. The crust of a neutron star (which
in our model is the only component since we do not consider the liquid interior) can
relax only through shearing motions as it spins down and so must be triaxial (Link,
Franco \& Epstein 1998; Franco, Link \& Epstein 2000). Adding the magnetic stresses,
which result from the multi-polar field near the surface would produce a very complicated
figure. It is, therefore, quite unlikely that the star is oblate axisymmetric ($e^2 = 0$)
or prolate axisymmetric ($e^2 \to \infty$) to a very high precision. On theoretical
grounds one might expect $e^2$ to be close to unity. Thus, we first take $e^2 \leq 2$
with a flat prior (Figs. \ref{pdf1A}-\ref{pdf5A}). However, we find that within this
range the pdf for $e^2$ is not confined. Because of that, we also consider a second case
where we allow for larger values of $e^2$ (Figs. \ref{pdf1B}-\ref{pdf5B}). The spindown
model we use is derived under the assumption that $e^2$ is not exceedingly large (see
Appendix C). Therefore, we take $e^2 \leq 4000$, which seems to encompass the regions of
interest, without violating our assumptions. Since the volume of integration is
considerably larger at large values of $e^2$ than at small values, taking a flat pdf over
$e^2$ in this case would greatly suppress the importance of small $e^2$. Therefore, we
need to incorporate a prior that is fair for both regimes: we take a flat prior over
$\ln(1+e^2)$, i.e. the prior for $e^2$ is $1/(1+e^2)$. For both $e^2 \leq 2$ and $e^2
\leq 4000$, we calculate pdfs for three different values of the maximum beam swing angle
constraint: $\Delta\vartheta_{max} = 1^\circ$, $3^\circ$ and $5^\circ$.

        \begin{figure}
        \centerline{\includegraphics[scale=0.34]{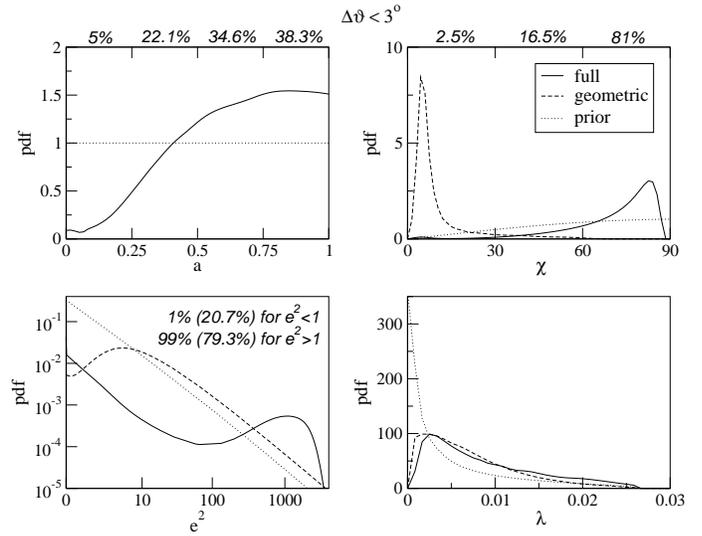}}
        \caption{Pdfs for beam swing angle variation less than $3^\circ$, for the same
        set of priors as in Fig. \ref{pdf1B}.}
        \label{pdf3B}
        \end{figure}

        \begin{figure}
        \centerline{\includegraphics[scale=0.34]{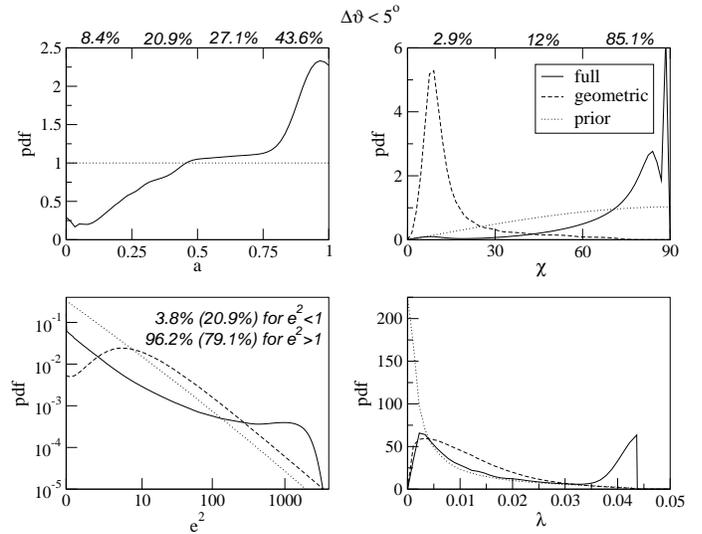}}
        \caption{Pdfs for beam swing angle variation less than $5^\circ$, for the same
        set of priors as in Fig. \ref{pdf1B}.}
        \label{pdf5B}
        \end{figure}

In Fig. \ref{plots} we show the data that we use, together with some sample models. The
parameters for these models are listed in Table 1. The best fit that we find is a purely
geometrical model, which has a very large beam swing angle that is, in fact, outside our
prior range (which was relaxed for determining an unconstrained, global ``favorite''
model). The axisymmetric model given here is similar to that of Link \& Epstein (2001),
except that the beam swing angle is constrained to be below $5^\circ$; in fact, as we
show in Appendix C, any axisymmetric model is assured to yield quite similar results even
when we introduce the additional torque parameter $a$.

Because of the large number of parameters, numerical integration for the pdfs is quite
time consuming. The figures presented here typically have a resolution of about 100
points per parameter or less. This means that fine structure in the pdfs may have been
missed. Nevertheless, the most notable structures in the pdfs are expected to remain.

        \begin{figure*}
        \centerline{\includegraphics[scale=0.7]{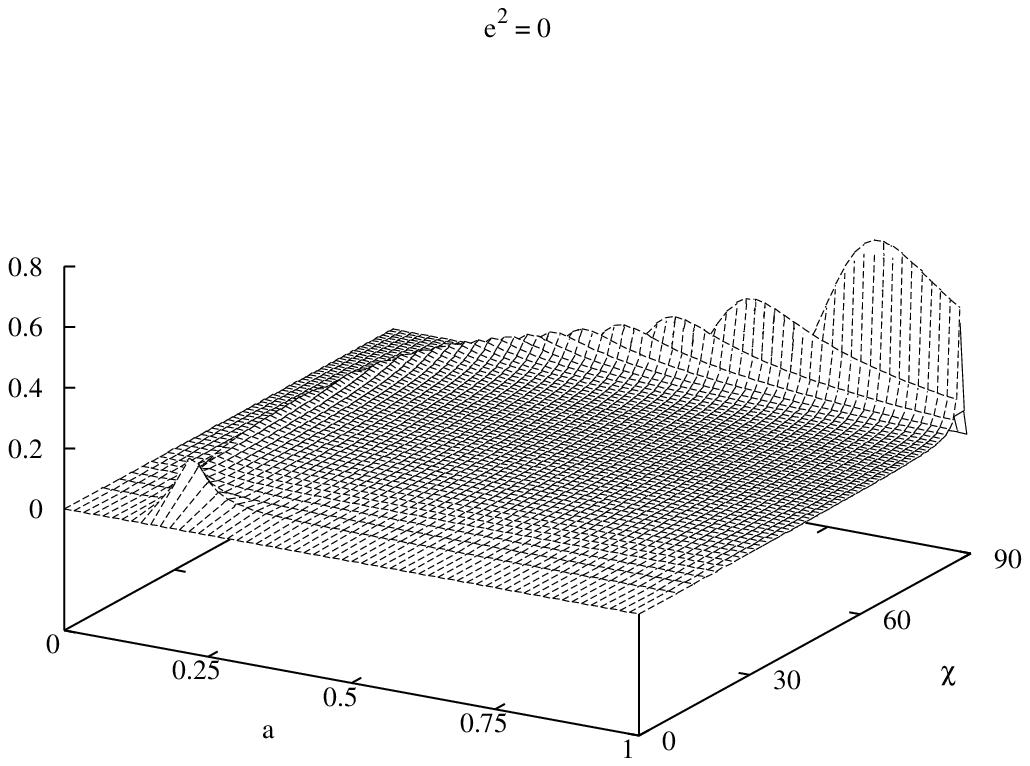}
        \includegraphics[scale=0.7]{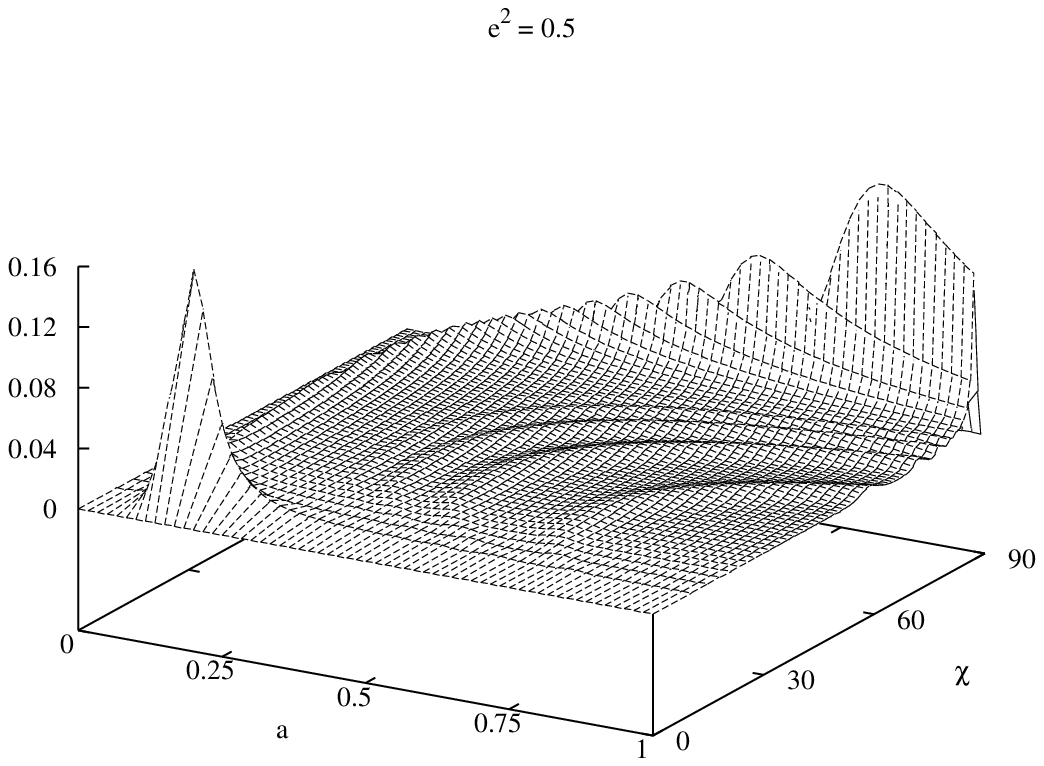}}
        \centerline{\includegraphics[scale=0.7]{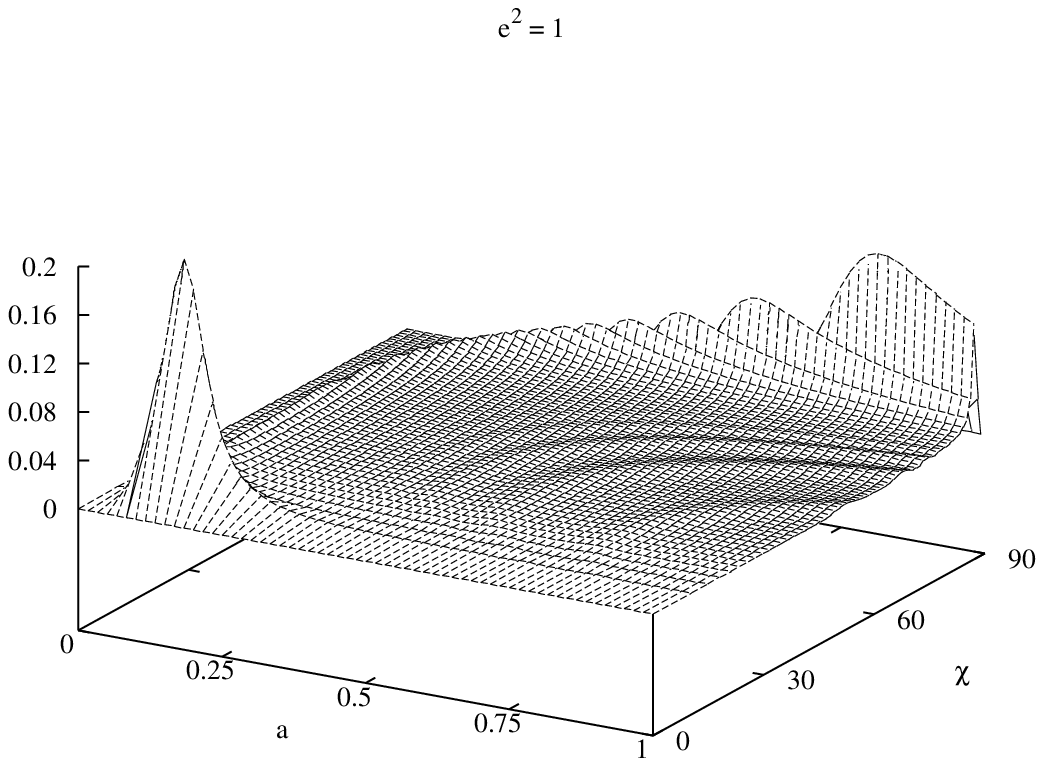}
        \includegraphics[scale=0.7]{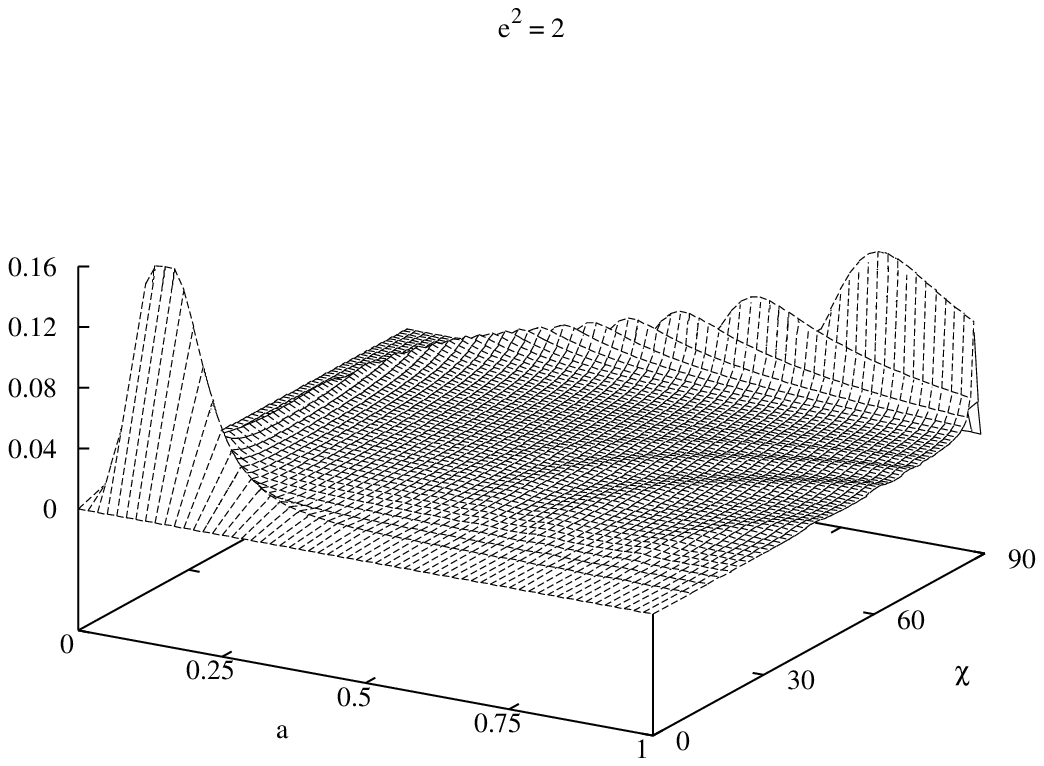}}
        \caption{Multivariate pdf surfaces for the full model at constant $e^2$ as
        functions of $a$ and $\chi$ for $\Delta\vartheta < 3^\circ$. The remaining
        parameters are integrated out, and normalization is carried out over all
        surfaces. The ripples are artifacts of integration.}
        \label{set3}
        \end{figure*}

In Figs. \ref{pdf1A}-\ref{pdf5B} we show the projected 1-D pdfs for each parameter,
computed by integrating the multidimensional pdf over all other parameters. The prior is
also a function of the entire set of parameters and is not separable for all except $a$.
We define the 1-D prior for a given parameter by integrating over the rest. A comparison
with the full 1-D pdf illustrates the importance of the period residuals in determining
the pdf. Keep in mind that both the prior and the posterior pdfs also include the beam
swing angle constraint. In these figures, the dotted lines are for the prior pdfs, the
dashed lines are for the geometric model alone (Eq. (\ref{residual})), and the solid
lines are for the geometric model and the spindown model (Eq. (\ref{dpsd})) both
combined.

We now discuss some of the main characteristics and implications of our analysis.

\emph{The torque parameter $a$:} For $e^2 \leq 2$ Figs. \ref{pdf1A}-\ref{pdf5A} show
considerable probability over the whole range of acceptable values, with a peak at low
$a$ that becomes more prominent as $\Delta\vartheta$ increases. Another lower and wider
peak appears for $\Delta\vartheta_{max} = 5^\circ$ at larger values of $a$ (Fig.
\ref{pdf5A}). Nevertheless, neither of the peaks is highly significant, because they do
not contain most of the probability. Therefore, we conclude that the data do not
constrain $a$ strongly, and it can be quite different from the vacuum spindown value $a
\equiv 1$. As larger values of $e^2$ are permitted, Figs. \ref{pdf1B}-\ref{pdf5B} show a
peak at $a \rightarrow 1$, but with a large tail extending over most of the parameter
space. With increasing values of $\Delta\vartheta$, lower $a$ values become likelier, but
most of the probability ($>90\%$) still lies at $a \geq 0.25$. Thus, the value of $a$ is
not well-determined, but there is evidence for an angle-dependent torque. The parameter
$a$ is truly a measure of the angle dependence of the spin-down torque; it does not
depend on the geometric effect at all.

\emph{The magnetic inclination $\chi$:} The axisymmetric model discussed by Link \&
Epstein (2001) requires $\chi$ to be extremely close to $90^\circ$. As discussed in
Appendix C, this is true even when we allow $a \neq 1$. For a triaxial model, we find the
range of acceptable $\chi$ values to be much larger. The geometric model has a peak at
small $\chi$, which moves on to higher values of $\chi$ and broadens with increasing beam
swing angle. This trend is still seen at $e^2 \leq 2$ when $a \neq 0$ is turned on. The
spindown produces a narrow but strong peak in the vicinity of $90^\circ$, which becomes
more pronounced as $\Delta\vartheta$ is allowed to be bigger. This peak corresponds to
the axisymmetric case, and implies that it requires larger $\Delta\vartheta$ values; in
fact, the model discussed by Link \& Epstein has $\Delta\vartheta \simeq 6.4^\circ$. For
$e^2 \leq 4000$ (Figs. \ref{pdf1B}-\ref{pdf5B}), the peak at large $\chi$ remains
apparent, though now it is quite broad. The inclusion of points beyond $e^2 \approx 2$
seems to favor a more important spindown contribution, and the geometric effect is
further suppressed. There is also a small cusp that appears in the pdf for
$\Delta\vartheta < 5^\circ$, at $\chi$ very near $90^\circ$, corresponding to the
axisymmetric case. Yet, this peak is quite narrow, and the vast majority of the
probability lies outside of it.

\emph{The triaxiality parameter $e^2$:} For $e^2 \leq 2$, the pdf looks quite similar to
the prior, implying that the data do not differentiate among values of $e^2$ (Figs.
\ref{pdf1A}-\ref{pdf5A}). There is a narrow sharp peak at $e^2 = 0$ for the full model,
corresponding to the \emph{oblate} axisymmetric case, but the probability enclosed within
the peak is very small. Note that, at the same time, the pdf for the geometric effect
alone almost vanishes, i.e. for the axisymmetric case, spindown is essential. When we
allow $e^2 > 2$, another peak appears in the pdf (Figs. \ref{pdf1B}-\ref{pdf5B}). At
these values of $e^2$ the star is once again almost axisymmetric, except that now $I_1 <
I_2 \simeq I_3$, i.e. the star is \emph{prolate}. Qualitatively $e^2 = 1$ separates
oblate and prolate shapes. Then, comparing the probabilities enclosed in the two regions,
$e^2 < 1$ and $e^2 > 1$ , we find that the prolate case contains, by far, most of the
probability, even though we have adopted a prior which somewhat disfavors large $e^2$
values. Note that as the beam swing angle constraint is relaxed, the probability becomes
quite evenly distributed over a wide region in $e^2$. This leads to the conclusion that
there are many triaxial models with a wide range of $e^2$ that can fit the data. In other
words, the data do not discriminate among values of $e^2$, especially when
$\Delta\vartheta$ is relatively large.

\emph{The wobble parameter $\lambda$:} For both $e^2 \leq 2$ and $e^2 \leq 4000$, the pdf
is contained in a region which seems to be mostly confined by the beam swing angle. The
constraint results in a dramatic cutoff at the high end of $\lambda$. Beyond that point,
we cannot find a value of $e^2$ for which $\Delta\vartheta$ will be smaller than the
constraint. Within that region, there is considerable probability distributed over the
whole range of $\lambda$; both models follow the same trend. We can conclude that the
oblate case favors somewhat larger values of $\lambda$, which produces a peak that is
partially visible for $\Delta\vartheta_{max} = 5^\circ$ (Fig. \ref{pdf5A}). The prolate
case, on the other hand, favors smaller values of $\lambda$, resulting in a second peak,
which appears when we allow $e^2$ to be large (Fig. \ref{pdf5B}), but is absent when
$e^2$ is confined to low values (Fig. \ref{pdf5A}).

Fig. \ref{set3} shows multivariate pdfs for the full model, plotted as surfaces at
constant $e^2$, as functions of $a$ and $\chi$. The remaining parameters ($\phi$,
$\lambda$ and $\tau_o$) are integrated out, and the pdfs are normalized over all
surfaces. (The special case $e^2 = 0$ can be done with considerably higher resolution,
because of the freedom of choice of $\phi$.) The ripples that are present are artifacts
of the integration and subside as the resolution is improved. The amplitude of the
ripples serves as an implicit way of evaluating the significance of the peaks in the pdf.
Relatively large amplitude implies that there are no significant peaks in the pdf,
meaning that no points or regions in the allowed parameter space are favored strongly.

As $\Delta\vartheta_{max}$ is increased, two regions acquire prominence: a very narrow
ridge at large $\chi$, which extends over a wide region in $a$ and includes the
axisymmetric case, and a smaller peak at small $\chi$ and small $a$. Keep in mind that
this second peak is further discriminated against by the prior, which is flat over
$\cos\chi$, i.e. there is a factor of $\sin\chi$ that also enters the pdf. Consequently,
when the one-dimensional pdfs are calculated, the second peak is considerably suppressed.

\section{Shape of the Pulse}

The pulse profile of PSR B1828--11 alternates between two different modes, one narrow and
the other broad, and has Fourier power at both 250 and 500 days (Stairs, Lyne \& Shemar
2000, Stairs et al. 2003). Stairs et al. (2003) describe how the pulse profile is
determined. During a particular observing session, 16 pulse averages may be either broad
or narrow, with a shape parameter $S$ defined to be the fraction of the mean pulse shape
for that session attributed to the narrow component. The shape parameter $S$ varies
systematically between $\approx 0$ (all wide) and $\approx 1$ (all narrow) over the
precession cycle, with a strong Fourier component at the ``first harmonic'' $1/250\,{\rm
days}$ of the ``fundamental'' precession frequency $1/500\,{\rm days}$ (Stairs, Lyne \&
Shemar 2000, Stairs et al. 2003). Link \& Epstein (2001) suggested that the emission beam
of the pulsar must have an hourglass shape in order for $S$ to exhibit substantial
variability on the 250 day timescale. (A similar elongated shape was inferred from
studies of the geodetic precession of PSR B1913+16 by Weisberg \& Taylor 2002.) However,
they did not address the issue of mode switching during individual observing sessions at
all. Here we present an alternative viewpoint centered around modeling profile mode
switching within individual observing sessions, and argue that it may be able to produce
some aspects of the required harmonic structure shown in Fig. \ref{data}.

        \begin{figure}
        \centerline{\includegraphics[angle=-90,scale=0.4]{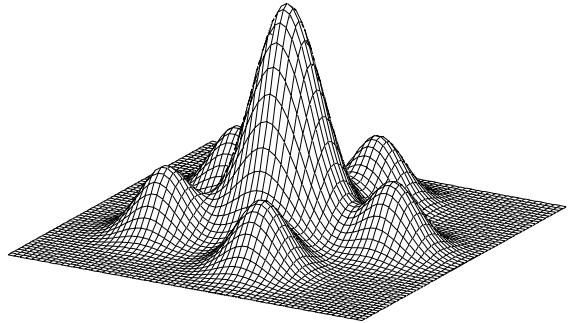}}
        \caption{Schematic of a pulse consisting of a bright core, and
        surrounded by smaller fainter conal blobs. Core emission is assured
        to be stationary, whereas conal emission could vary as the emitting blobs
        circulate about the beam axis.}
        \label{shape}
        \end{figure}

The basic {\it geometrical} picture is shown in Fig. \ref{shape}. We attribute the narrow
component of the pulse to {\it core} emission centered around the beam axis.  The broader
profile is a superposition of core and conal emission. Thus, in the parlance of Rankin
(1990, 1993), the pulsar alternates between presenting a core single (${\rm S_t}$) and
triple (T) pulse profile. The relatively young spindown age of PSR B1828--11 (about 0.11
Myr) and its large value of $B_{12}/P^2$ are consistent with this categorization.
However, the apparent pulse width in the narrow state appears to be anomalous: Rankin
(1990) finds a FWHM pulse width $\wc=2.45^\circ\, P^{-1/2}/\sin\chi$, where $\chi$ is the
angle between a pulsar's spin and magnetic axes. For PSR B1828--11, the FWHM of the
narrow state is about $2.3^\circ$, as opposed to $\wc\approx 3.85^\circ/\sin\chi$ from
Rankin's (1990) formula. We note that the bounding relationship
$\wc\geq2.45^\circ/P^{-1/2}$ was derived from a set of ``interpulsars'' thought to be
nearly orthogonal rotators, so we would have expected Rankin's (1990) formula to work
especially well if $\chi$ is near $90^\circ$. as was suggested by Link \& Epstein (2001);
on the other hand, the discrepancy is also smallest for $\chi\approx 90^\circ$, which may
be circumstantial evidence that PSR B1828--11 really is nearly an orthogonal rotator.
(Moreover, the frequency dependence of the core width is relatively weak at high
frequencies, so that the fact that the Rankin [1990] formula is for 1 GHz emission,
whereas the Stairs et al. [2003] observations were at 1.6 or 1.7 GHz is not responsible
for the discrepancy.) We note that there are other exceptions to Rankin's (1990) bound,
but not many (see e.g. Fig. 23b in Graham-Smith [2003], adapted from Gould [1994]); given
uncertainties in $\chi$ there may be other pulsars with $\wc>2.45^\circ/\sqrt{P}$ but
also $\wc<2.45^\circ/\sqrt{P}\sin\chi$.

If pulsar core emission intensity were Gaussian, we would expect an observed intensity of
the form,
        \begin{equation}
        \icore=\icore(0)\exp\left[-{\beta^2\over 2\rho_1^2} -
        {(\phi\sin\alpha)^2\over 2\rho_2^2}\right]~,
        \label{gbeam}
        \end{equation}
where $\beta$ is the impact parameter of the observer's line of sight relative to the
beam axis, $\phi$ is pulse phase (centered on epoch of closest passage relative to the
axis), and $\alpha=\chi+\beta$ is the angle between the line of sight to the observer and
the stellar spin axis. Here $\rho_1$ and $\rho_2$ define the extent and the shape of the
beam, which would be elliptical when they are not equal. This formula assumes that
$\beta\ll\chi$, and that emission is strongly beamed along magnetic field lines, but does
not presume that the emission pattern is circularly symmetric with respect to the beam
axis. The two directions 1 and 2 are relative to a coordinate system in which $\he_3=\hb$
coincides with the magnetic moment of the star, which is assumed to be the beam axis,
$\he_2=\hL\crossprod\hb/\vert\hL\crossprod\hb\vert$, and
$\he_1=(\hb\dotprod\hL\hb-\hL)/\vert\hL\crossprod\hb\vert$. For a Gaussian beam, Eq.
(\ref{gbeam}) shows that the core component width is independent of the impact angle
$\beta$, although the peak intensity is not (e.g. Rankin 1990). However, this is a unique
property of a Gaussian profile. We can well imagine that the emission cuts off sharply
(even discontinuously) for sufficiently large $\beta$, in which case the core width could
be narrower than normal. Because the peak core intensity would also be lower in such
cases, it would be harder to detect, which may account for the rarity of exceptions to
Rankin's (1990) bound.

A sharper cutoff to the core emission beam would not only allow narrower core pulse
profiles, but would also introduce $\beta$ dependence into the width. As a simple
example, suppose that the beam profile is,
        \beq
        \icore&=&\icore(0)\exp\left(-{u\over 2}-{\kappa u^2\over 4}\right)
        \nonumber\\
        u&=&{\beta^2\over\rho_1^2}+{(\phi\sin\alpha)^2\over \rho_2^2}~,
        \enq
i.e. still a self-similar function but with a sharper cutoff than a Gaussian profile. The
peak intensity is at $\phi=0$, where $u=\umin=\beta^2/\rho_1^2$; the FWHM is at phases
$\pm\phi_{1/2}$, where
        \beq
        {\phi_{1/2}\sin\alpha\over\rho_2} \!\!\! &=& \!\!\!
        \left[\sqrt{\left(\umin+{1\over \kappa}\right)^2+{4\ln 2\over \kappa}}
        -\left(\umin+{1\over \kappa}\right)\right]^{1/2}\nonumber\\
        &\approx& \!\!\! \sqrt{{2\ln 2\over 1 + \kappa\umin}}=
        \sqrt{{2\ln 2\over 1 + \kappa\beta^2/\rho_1^2}}
        \enq
where the approximation is valid for small values of $\kappa$, irrespective of
$\kappa\umin$. The cutoff becomes important once $\kappa\umin\sim 1$ i.e. for
$\beta\ga\rho_1/\sqrt{\kappa}$. The core width decreases with increasing $\beta$, as does
the peak intensity observed from the core.

As we mentioned above, we ascribe the broader pulse profile state to a superposition of
core and conal components. In keeping with the schematic Fig. \ref{shape}, we assume that
the conal emitting pattern is patchy and, as we discuss further below, probably only
stationary in the mean. Consider an individual conal emitting region (hereafter ``blob'')
$i$; we assume that it is centered at $(x_i^1,x_i^2)=\rhoi(\cosi, \,\sini)$. The emission
pattern of blob $i$ may be anisotropic in a complicated fashion, with possible preferred
directions not only along the $\he_{1,2}$ axes, but also along and perpendicular to
$\hx_i=(\cosi,\,\sini)$. The observer sees an intensity that is a function of the two
variables $\beta-\rhoi\cosi$ and $-(\phi+\phi_d) \sin\alpha-\rhoi\sini$, where the phase
delay is $\phi_d= h_i/cP_\star\approx 1^\circ(h_i/340\,{\rm km})$, where $h_i$ is the
height of blob $i$ above the core emitting region. The peak value of the intensity of
radiation seen from any given blob is only a function of $\vert\beta-\rhoi\cosi\vert$,
though, and we assume that blob $i$ is detectable provided that
        \begin{equation}
        \vert\beta-\rhoi\cosi\vert\leq\delta_i~,
        \end{equation}
where $\delta_i$ may depend on $\sigma_i$. Thus, the detectability of an individual blob
varies through the precession cycle, and the probability of seeing any blobs at all also
varies, thus affecting the observed beam width.

        \begin{figure}
        \centerline{\includegraphics[scale=0.35]{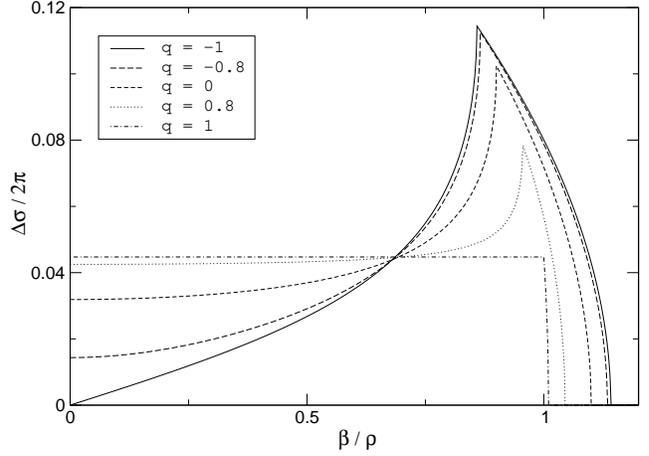}}
        \caption{$\Dsig/2\pi$ vs. $\beta$ for various
        $q$ and $\delta/\rho=0.1$.}
        \label{fig2}
        \end{figure}

        \begin{figure}
        \centerline{\includegraphics[scale=0.35]{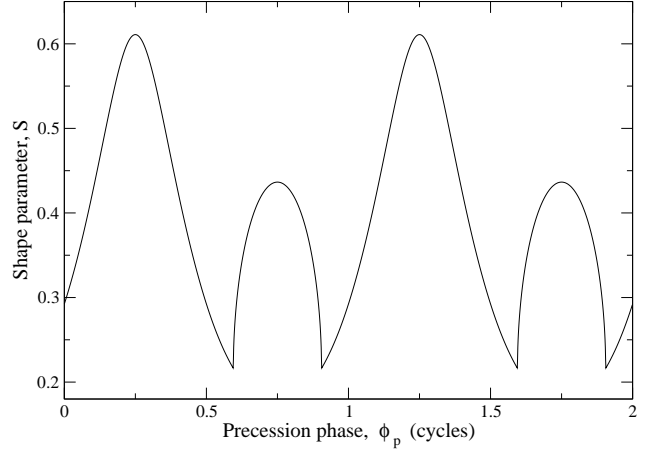}}
        \caption{``Shape parameter'' vs. precession phase for a
        model with $\delta/\rho=0.1$, $q=-0.8$, $N=6$, and impact parameter
        $\beta/\rho=0.95+0.15\sin\phi_p$.}
        \label{fig3}
        \end{figure}

The problem of modeling the detectability of a given blob can be quite complex. To
illustrate, suppose that each blob is at $\rhoi=\rho$, and has the same anisotropic
shape. Simplify even more by assuming that the emission profiles of the blobs have a
characteristic length $\deltar$ along the (radial) direction from the beam axis to its
center, and a different length $\deltat$ in the direction tangential to it. Then we can
detect the blob if,
        \beq
        (\beta-\rho\cosi)^2 \leq \deltat^2\sinis+\deltar^2\cosis
        \label{deltaview}
        \enq
For convenience, we define $\delta^2={1\over 2}(\deltat^2+\deltar^2)$ and
$q\delta^2={1\over 2}(\deltat^2-\deltar^2)$. Note that $q$ may be positive or negative,
and $\vert q\vert\leq 1$; $q=0$ for a circular beam. Since we expect that in general
$\delta\ll\rho$, and except for special values, $\delta\ll\beta$, we only expect blobs
within a small range $\Dsig(\beta)$ to be visible.

Fig. \ref{fig2} shows the result of solving Eq. (\ref{deltaview}) for the range of
observable $\Dsig/2\pi$ as a function of impact parameter $\beta$. (The solutions were
not extended beyond $\beta/\rho=1+\delta\sqrt{1-q}$, where $\Dsig\equiv 0$.) The results
exhibit complicated behavior even in this simple model. Given $N$ conal blobs, the
probability of seeing the broader pulse profile is large when $2\times
N\Delta\sigma/2\pi\ga 1$, and is small when $N\Delta\sigma/\pi\la 1$. (The factor of two
is because the observer's line of sight crosses the cone twice.) Thus, we may expect the
shape parameter $S$ to be small for impact parameters where $\Dsig/2\pi$ is large, and
vice-versa; during a precession cycle, both regimes may be sampled.

Fig. \ref{fig3} shows an example of how the probability of seeing only the narrow pulse
would vary with precession phase in this model; this example captures the main features
of the observed beam width variations shown in Fig. \ref{data}.  For constructing the
figure we adopted $q=-0.8$, $\delta=0.1\rho$, and assumed a sinusoidal variation of the
impact parameter with precession phase:
        \begin{equation}
        \beta(\phi_p)=\beta_0+\beta_1\sin\phi_p
        \end{equation}
with $\beta_0=0.95\rho$ and $\beta_1=0.15\rho$ assumed for graphical purposes. The
``shape parameter'' is taken to be $S=(1-\Dsig/\pi)^N$ i.e. the probability that no blobs
are detected; Fig. \ref{fig3} assumes $N=6$. Clearly, $S$ varies periodically but not
sinusoidally in this model, and also varies substantially in half a precession cycle.
This distinctive ``doubly periodic'' variation is only seen if the observer's line of
sight crosses near $\beta=\rho$. This is consistent with our earlier discussion of core
widths if the core emission is still visible but starting to cut off at such impact
angles. Presumably, the peak intensity of the core emission must also far exceed that of
any conal blob for this model to be viable; there are some indications that conal
emission becomes more prominent as pulsars age (e.g. Rankin 1990). Although the range of
variation of $S$ in this example is smaller than in PSR B1828--11, extensions of the
model, such as different assumptions about the conal emission (e.g. an hourglass-shaped
cone as in Link \& Epstein [2001], or a more complicated version of blob anisotropy) may
possibly yield a better account of the data.

If the conal blobs were stationary in the rotating frame, then the observer would see
pulse profile variations as a function of precession phase, but would not see any
variations at a given precession phase. However, it is likely that the conal blobs are
not at fixed positions but rather circulate around the cone in a rotating carousel
(Deshpande \& Rankin 1999, 2001; see Rankin \& Wright 2003 for a review). In this
picture, which has empirical support (Deshpande \& Rankin 1999, 2001), conal emission is
from beams that circulate with a frequency $\Omega_d=f_d\Omega_\star$ relative to a
reference frame rotaing with the star; the circulation is probably the consequence of
$\Evec\crossprod\Bvec$ drift (e.g. Ruderman \& Sutherland 1975, Gil, Melikidze \& Geppert
2003, Wright 2003), and $f_d\la 0.1$ is a reasonable value. By contrast, the core
emission is stationary, and from a much lower altitude than the conal emission (possibly
from near the polar cap). In this picture, during a given observing session the core
component is always visible, but the conal component fluctuates as emitting blobs pop in
and out of the observer's line of sight periodically. The probability that the observer
sees conal emission at all varies systematically during the precession cycle, and is
fixed during any observing session lasting a day or so (i.e. far less than the precession
period).

If this model is correct at least in a broad outline, then the total beam swing during a
precession cycle is $\leq 1-3^\circ$, given expected core radii (Rankin 1993). Larger
beam swings could be accommodated by a more complex model for the pulse shape (eg. the
hourglass shape of Link \& Epstein 2001).

\section{Conclusion}
We find a wide range of triaxial models that may explain the period residuals of PSR
B1828--11 in terms of precession, even under the stringent constraints we have imposed.
We find many fits that are as good as, or better than the axisymmetric model considered
before (Link \& Epstein 2001). In general, fits improve with larger beam swing angle
variations ($\Delta\vartheta$), but if we assume that the pulse is confined to a region a
few degrees in size, we have to rule them out. Prolate and oblate axisymmetric models
seem to be favored, especially for small $\Delta\vartheta$, but that is not sufficient to
rule out other triaxial models. Both the geometric and spindown effects contribute to the
fits. Oddly, if we relax our beam swing constraint completely, the data prefer a best fit
that has $a = 0$ (no spindown contribution), but $\Delta\vartheta$ for that model is
unreasonably large (Fig. \ref{plots} and Table 1), so it is merely an unphysical
curiosity.

In the oblate axisymmetric model, spindown is the dominant effect, but it requires
parameters (in particular, a large $\chi$ value) that could be expected to produce an
interpulse, which is not seen in PSR B1828--11. If we were to impose the absence of an
interpulse as a constraint, some of the models we have permitted in our analysis would be
excluded, particularly those at large $\chi$. Conceivably, the magnetic field and core
beam structure of PSR B1828--11 are sufficiently complex that an interpulse would be
absent even at $\chi\to 90^\circ$. We note that our model for shape variations suggests
that we are only viewing the outskirts of the core emission in the component we detect,
which may enhance the probability that emission from the opposite pole is undetectable.
Thus, we do not impose the absence of an interpulse as a constraint on our analysis.

Our models do not require $a = 1$, so substantial deviations from the vacuum spindown
formula are allowed. In fact, rather small values of $a$ are permitted for an oblate star
($e^2 < 1$). However, for a prolate star ($e^2 > 1$) we find that larger values are
favored ($a>0.25$), thus providing evidence for an angle-dependent torque. {\em To our
knowledge, our analysis provides the first evidence that pulsars are spun down by a
torque that depends on the angle between the magnetic moment and the instantaneous
angular velocity}.

The magnetic obliquity, $\chi$ is no longer required to be extremely close to $90^\circ$,
and we find $\lambda$ (which is related to the wobble angle) to be restricted mainly by
the beam swing angle. Two peaks in $e^2$ are prominent, corresponding to the oblate
axisymmetric ($e^2 = 0$) and prolate nearly axisymmetric (large $e^2$) cases. These are
especially evident for small beam swing angle variations. For larger beam swing
variations, the data do not discriminate among values of $e^2$ very much.

In summary, our precession model fits the data equally well for a broad range of
parameters. We cannot constrain the shape of the star, but we {\em do} find evidence for
angle-dependent spin-down torque. Overall, the ability of our physically-motivated model
to account for the principal features of PSR B1828--11's timing data without special
choices of the parameters reinforces the idea that the pulsar is precessing. In
particular, we have shown that the data can be fit without resorting to a nearly
orthogonal rotator with a vacuum-like dipole torque as Link \& Epstein (2001) did in
their preliminary work. Though we cannot strongly constrain the angular dependence of the
spin-down torque with the present data, the potential remains for learning more about
this important aspect of the neutron star magnetosphere from future observations. The
parameters of our model are not very tightly constrained. Two possible reasons are that
we have only three cycles of data, and that the data have a large degree of intrinsic
scatter that our simple model cannot account for, creating wide pdfs. It will be
interesting to see if the parameters can be more tightly constrained as more data become
available over the next decade.

Our analysis does not employ the data on the shape parameter variations, because
constructing a comprehensive mathematical description would require a reliable model for
the pulsar beam. Although we do not possess such an accurate model, we offer an
explanation using a compound pulse structure, with core and cone components.

Precession has interesting implications for pulsar observations, which so far
have not been widely discussed. One immediate, and very obvious effect would be
\emph{disappearing pulsars}, i.e. pulsars that due to precession, would at some
point leave the line of sight of the observer, but excluding other effects, would
eventually come back. The timescale of such changes could be months to years.

\section*{Acknowledgments}

We thank Ingrid Stairs for providing us with the data and for valuable
discussion. This research is supported in part by NSF AST-0307273 (Cornell
University), NSF AST-0098728 (Montana State University) and IGPP 1222R from LANL.

\appendix
\onecolumn

\section{Period Residuals for a Triaxial Rigid Star in the Absence of External
Torques: The Geometric Contribution}

Euler's equation for a freely precessing rigid body is,
        \beq
        &&\frac{d\m{L}}{dt} + \m{\Omega}\times\m{L} = 0
        \enq
and can be solved analytically in terms of Jacobian elliptic functions (see Landau \&
Lifshitz, 1976). Using the principal axes ($I_1 \leq I_2 \leq I_3$) as the basis for the
body (rotating) frame, we can express the components of the angular momentum unit vector
$\hL$ as,
        \beq
        &&L_1 = - \Lambda_1 \cn (\tau,k^2) \mbox{ , }
        \Lambda_1 = \sqrt{\frac{I_1(2EI_3-L^2)}{L^2(I_3-I_1)}}
        \nonumber \\
        &&L_2 = - \Lambda_2 \sn (\tau,k^2) \mbox{ , }
        \Lambda_2 = \sqrt{\frac{I_2(2EI_3-L^2)}{L^2(I_3-I_2)}} = \Lambda_1\sqrt{1+e^2}
        \label{landau} \\
        &&L_3 = \Lambda_3 \dn (\tau,k^2) \mbox{ , }
        \Lambda_3 = \sqrt{\frac{I_3(L^2-2EI_1)}{L^2(I_3-I_1)}} = \sqrt{1 - \Lambda_1^2}
        \nonumber
        \enq
Here, the argument of the elliptic functions is,
        \beq
        &&\tau = t \omega_p \hspace{0.6cm} \mbox{where} \hspace{0.6cm}
        \omega_p = \sqrt{\frac{(I_3-I_2)(L^2-2EI_1)}{I_1I_2I_3}} =
        \frac{\epsilon L\Lambda_3}{I_3\sqrt{1+e^2}}
        \label{tau}
        \enq
and, the parameter of the elliptic functions is,
        \beq
        &&k^2 = \frac{(I_2-I_1)(2EI_3-L^2)}{(I_3-I_2)(L^2-2EI_1)}
        = \frac{e^2\Lambda_1^2}{\Lambda_3^2} = e^2 \lambda^2
        \label{lambda}
        \enq
where, we make use of the following auxiliary definitions,
        \beq
        &&\epsilon = (I_3 - I_1)/ I_1 \hspace{0.6cm} \mbox{ and } \hspace{0.6cm}
        e^2 = \frac{I_3(I_2-I_1)}{I_1(I_3-I_2)}
        \label{e2}
        \enq
The minus signs that we have explicitly included in our definitions are due to our
choice of the initial orientation of axes.

Note that $\omega_p$ is not the precession frequency, since the elliptic
functions do not have a period of $2\pi$, or more precisely, $\omega_p$ is not
the time derivative of the angular displacement. Instead, the precession
frequency is given through,
        \beq
        &&\Omega_p = \frac{2\pi}{P_p} = \frac{\omega_p \pi}{\tilde{\pi}}
        \enq
where 2$\tilde{\pi}$ is the period of the elliptic functions and can be calculated
using the Legendre elliptic integral of the first kind, $\tau = F(\phi,k)$ where
$\sin\phi = \sn\tau$,
        \beq
        &&\tilde\pi/2 = F(\pi/2, k)
        \enq

The values of the parameter $k^2$ are unrestricted, though different regimes require
careful handling. $k^2 < 1$ corresponds to precession around the body $z$ axis; $k^2 = 1$
corresponds to the unstable trajectories of the angular momentum, which decay
exponentially towards the intermediate axis, $y$; and $k^2 > 1$ is precession around the
$x$ axis (see Binet ellipsoid). For now we will confine ourselves to the first case, and
the other two will be left for a later section.

\begin{figure}
        \centerline{\includegraphics[scale=0.4]{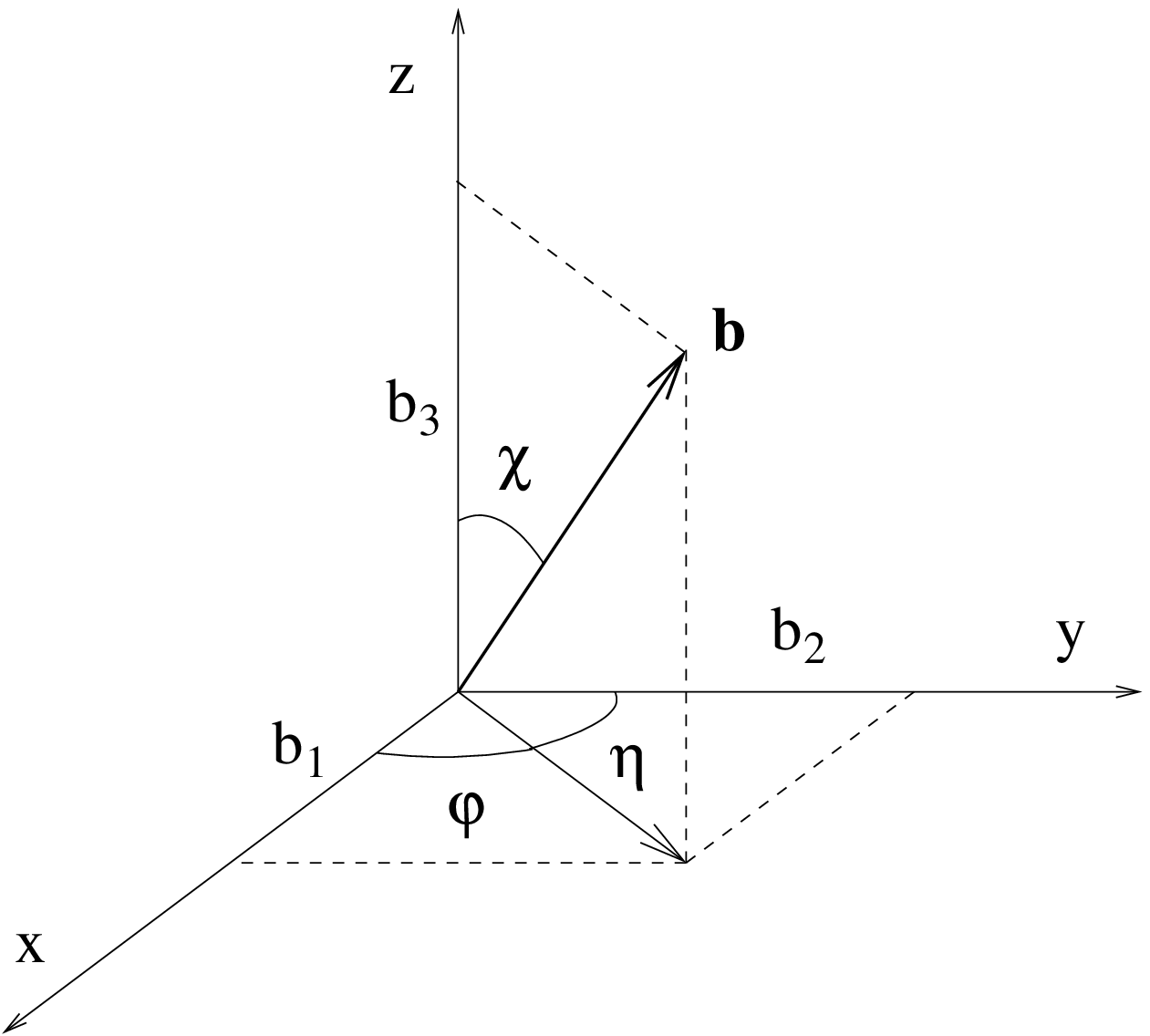}}
        \caption{Orientation of the magnetic axis in the body frame. We assume that the pulses
        are also emitted along the same axis.}
        \label{pulsaxis}
        \end{figure}

We are interested in an isolated neutron star, and we want to determine the time of
arrival (TOA) of pulses (produced along the magnetic axis) for an inertial observer. Let
the inertial $z$ axis be along the angular momentum vector, which remains constant; and
let the inertial $x$ axis be defined by the orientation of the observer, whom we choose
to locate in the first quadrant of the inertial $xz$ plane. Let a unit vector $\hb$
denote the orientation of the magnetic axis, and $b_i$ be the rotating frame components.
Then, whenever the inertial $y$ component (which we choose to denote by $b_y$) vanishes,
while the inertial $x$ component ($b_x$) is positive, we get a pulse. The two frames are
related through a rotation matrix constructed from the Euler angles $\theta, \psi$ and
$\phi$ (see Goldstein 1980), whence the two conditions can be expressed as,
        \beq
        &&b_y \ = \ b_1 (\cos\psi\sin\phi + \cos\theta\cos\phi\sin\psi)
        + b_2 ( -\sin\psi\sin\phi + \cos\theta\cos\phi\cos\psi)
        \ - \ b_3 \sin\theta\cos\phi \ = \ 0
        \label{pulsey}
        \enq
and,
        \beq
        &&b_x \ = \ b_1 (\cos\psi\cos\phi - \cos\theta\sin\phi\sin\psi)
        - b_2 (\sin\psi\cos\phi + \cos\theta\sin\phi\cos\psi)
        \ + \ b_3 \sin\theta\sin\phi \ > \ 0
        \label{pulsex}
        \enq
Using the solution for the angular momentum (Eq. (\ref{landau})), the Euler
angles are given through,
        \beq
        &&\cos\theta = \Lambda_3 \dn \tau \mbox{ , } \hspace{0.3cm}
        \sin\theta = \Lambda_1 \sqrt{1+e^2\sn^2\tau} \mbox{ , } \hspace{0.3cm}
        \cos\psi = - \frac{\sqrt{1+e^2} \sn\tau}{\sqrt{1+e^2\sn^2\tau}}
         \mbox{ , } \hspace{0.3cm}
        \sin\psi = - \frac{\cn\tau}{\sqrt{1+e^2\sn^2\tau}} \hspace{0.3cm} \mbox{ and }
        \label{euler} \\
        &&\frac{d\phi}{dt} = \frac{L}{I_3}\left( 1 +
        \frac{\eps}{1+e^2 \sn^2 \tau}\right) \nonumber
        \enq
The last equation can be written as,
        \beq
        &&\phi (t) = \phi_o + \frac{L}{I_3}t + \frac{\sqrt{1+e^2}}{\Lambda_3}
        \int_0^\tau \frac{d\tau}{1+e^2\sn^2\tau}
        \label{intphi}
        \enq
Eq. (\ref{pulsey}) also implies,
        \beq
        &&\tan\phi \ = \ \frac{N}{D} \nonumber \\
        &&N \ = \ b_3\Lambda_1 \left(1+e^2\sn^2\tau\right) \ + \ b_2\Lambda_3\sn\tau\dn\tau
        \sqrt{1+e^2} + \ b_1\Lambda_3\cn\tau\dn\tau \label{tanphi} \\
        &&D \ = \ b_2\cn\tau - b_1\sn\tau\sqrt{1+e^2}\nonumber
        \enq
Pulses are seen when Eqs. (\ref{intphi}) and (\ref{tanphi}) are both satisfied.
In the absence of precession ($\Lambda_1 = 0$ and $k^2 = 0$) the solution of Eq.
(\ref{tanphi}) for $\phi$ is simply given through,
        \beq
        &&\phi = 2\pi n + \eta + \tan^{-1}(\sqrt{1+e^2} \ \tan\tau)
        \enq
where $\eta = \pi/2 - \varphi$ (Fig. \ref{pulsaxis}); and we can further restrict it to
lie anywhere between 0 and $2\pi$. Note that we have implicitly included the second
requirement, Eq. (\ref{pulsex}), by skipping every other possible solution for $\phi$.
This is a non-trivial assumption, and would break down if the magnetic axis $\m{b}$
happens to lie between $\Momega$ and $\m{L}$. However, for a pulsar these two vectors are
very nearly aligned since $\epsilon$ is extremely small. Therefore, we do not need to
worry about such a case.

We express the general solution for $\phi$ as,
        \beq
        &&\phi = 2 \pi n + \eta + \zeta
        \label{phitwo}
        \enq
where $\zeta$ is confined to lie within a period of tangent (i.e. $\pi$). Using
$\tan\eta = b_1/b_2$ one can show that,
        \beq
        &&\tan\zeta = \frac{Nb_2 - Db_1}{Nb_1 + Db_2}
        \label{delta}
        \enq
We now have two equations for $\phi$ (Eqs. (\ref{intphi}) and (\ref{phitwo}))
which we can combine to get the times of arrival of pulses,
        \beq
        &&\frac{L}{I_3} \ t_n = 2\pi n + \zeta_n - \zeta_o -
        \frac{\sqrt{1+e^2}}{\Lambda_3}
        \int_0^{\tau_n} \frac{d\tau}{1+e^2\sn^2\tau}
        \label{timen}
        \enq
$\zeta_o$ (for $\tau=0$) appears as a consequence of the fact that $\phi_o = \eta
+ \zeta_o$. We have,
        \beq
        &&\tan\zeta_o = \frac{b_2b_3\Lambda_1 - b_1b_2 (1 - \Lambda_3)}
        {b_1b_3\Lambda_1 + b_2^2 + b_1^2\Lambda_3}
        \enq
Note that in the absence of precession (i.e. when $\Lambda_1 = 0$) the times of
arrival reduce to the form,
        \beq
        &&t_n = \frac{2\pi I_3 n}{L}
        \label{timenp}
        \enq
which is the solution for pure rotation.

The period (between two pulses) is given as,
        \beq
        &&P_n = t_n - t_{n-1}
        \enq
whence,
        \beq
        &&\frac{L}{I_3} P_n - 2\pi \ = \ \frac{L}{I_3} \Delta P_n
        \ = \ \zeta_n - \zeta_{n-1} - \frac{\sqrt{1+e^2}}{\Lambda_3}
        \int_{\tau_{n-1}}^{\tau_n} \frac{d\tau}{1+e^2\sn^2\tau}
        \enq
If the precession period is much longer than the pulse period (as is the case for
a neutron star) we can approximate the differences by derivatives,
        \beq
        &&\frac{L}{I_3} \Delta P_n
        = \frac{d\zeta_n}{dn} -
        \frac{\sqrt{1+e^2}/\Lambda_3}{1+e^2\sn^2\tau_n} \ \frac{d\tau_n}{dn}
        = \left[ \frac{d\zeta_n}{d\tau_n} -
        \frac{\sqrt{1+e^2}/\Lambda_3}{1+e^2\sn^2\tau_n}
        \right] \frac{d\tau_n}{dn}
        \enq
We will find it convenient to define the expression inside the parentheses as a
new function,
        \beq
        &&f_n = \frac{d\zeta_n}{d\tau_n} -
        \frac{\sqrt{1+e^2}/\Lambda_3}{1+e^2\sn^2\tau_n}
        \label{func}
        \enq
The derivative $d\zeta/d\tau$ is given through (from Eq. (\ref{delta})),
        \beq
        &&\frac{d\zeta}{d\tau} = \frac{DN^\prime - ND^\prime}{N^2 + D^2}
        \label{ddelta}
        \enq
where, from Eq. (\ref{tanphi}),
        \beq
        &&\frac{d N}{d\tau} \ = \ 2b_3\Lambda_1 e^2 \sn\tau\cn\tau\dn\tau
        + b_2\Lambda_3\cn\tau (\dn^2\tau - k^2\sn^2\tau)\sqrt{1+e^2}
        - b_1\Lambda_3 \sn\tau(\dn^2\tau + k^2\cn^2\tau) \nonumber \\
        &&\frac{d D}{d\tau} \ = \ - b_2\sn\tau\dn\tau - b_1
        \cn\tau\dn\tau \sqrt{1+e^2} \nonumber
        \enq

To evaluate $d\tau_n/dn$ we will make use of the time of arrival equation, Eq.
(\ref{timen}). Taking the derivative of both sides with respect to $n$, we get,
        \beq
        &&\frac{L}{I_3}\frac{dt_n}{d n} = \frac{L}{I_3 \omega_p} \frac{d\tau_n}{d n}
        = 2\pi + f_n \frac{d\tau_n}{d n}
        \hspace{0.6cm} \mbox{ so that } \hspace{0.6cm}
        \frac{d\tau_n}{d n} = \frac{2\pi\varpi_p}{1 - \varpi_p f_n}
        \enq
where we have defined a new dimensionless quantity,
        \beq
        &&\varpi_p = \frac{I_3\omega_p}{L} = \frac{\epsilon\Lambda_3}{\sqrt{1+e^2}}
        \enq
The pulse period will be given through,
        \beq
        &&P_n = t_n - t_{n-1} \sim \frac{dt_n}{d n}
        = \frac{1}{\omega_p} \frac{d\tau_n}{d n} = \frac{P_\star}{1 - \varpi_p f_n}
        \label{period}
        \enq
whence the period residuals can be found to be,
        \beq
        &&\frac{\Delta P_n}{P_\star} \equiv \frac{P_n}{P_\star} - 1 =
        \frac{\varpi_p f_n}{1 - \varpi_p f_n} \sim \varpi_p f_n
        \label{residual}
        \enq
where $P_\star = 2\pi I_3/ L$ is the rotation period of the star, and the last
approximation results from our anticipation that $\epsilon$ will be sufficiently
small. Indeed, from the above definitions we get, for small $k^2$,
        \beq
        &&\frac{\eps}{\sqrt{1 + e^2}} \sim \frac{P_\star}{P_p} \sim 3.2\times 10^{-8}
        \frac{P_\star \mbox{ (sec)}}{P_p \mbox{ (yrs)}}
        \enq

The coefficient of the function $f_n$ in Eq. (\ref{residual}) is,
        \beq
        &&B = P_\star \varpi_p \hspace{1cm} \mbox{where} \hspace{1cm}
        \varpi_p = \frac{I_3\omega_p}{L} = \frac{2\tilde\pi I_3}{LP_p} =
        \frac{\tilde\pi P_\star}{\pi P_p}
        \enq
in other words,
        \beq
        &&B = \frac{\tilde\pi P_\star}{\pi P_p}^2 = \frac{\tilde\pi B_o}{\pi}
        \enq
For PSR B1828--11 the rotation period is 405.04 ms, and the precession period is
about 511 days, so that $B_o \simeq 3.8$ ns.

\subsection{The Axisymmetric Body}

For an axisymmetric body we have $e^2 = k^2 = 0$. We can also set $b_2 = 0$ which
is equivalent to introducing some initial phase shift in the definition of
$\tau$. We thus get, after some rearrangement,
        \beq
        &&f_n = \frac{d\zeta}{d\tau} - \frac{1}{\Lambda_3}
        = -\frac{\Lambda_1}{\Lambda_3}
        \frac{b_1b_3\Lambda_3\cos\tau + b_1^2\Lambda_1\sin^2\tau + b_3^2\Lambda_1}
        {(b_3\Lambda_1 + b_1\Lambda_3\cos\tau)^2 + b_1^2\sin^2\tau}
        \enq

Let's now assume that the angle $\theta$ between the symmetry axis and the
angular momentum is small, i.e. $\Lambda_1 \sim \theta$ and $1-\Lambda_3 \sim
\theta^2/2$, and working to second order compute the period residuals,
        \beq
        &&f_n \approx - \theta \left(\frac{b_3}{b_1}\right)\cos\tau - \frac{\theta}{2}^2
        + \theta^2 \left[\frac{1}{2} + \left(\frac{b_3}{b_1}\right)^2\right] \cos 2\tau
        \enq
Also let $b_3 = \cos\chi$. Then,
        \beq
        &&\Delta P_n / P_\star \approx \varpi_p f_n \approx - \theta \epsilon \cot\chi
        \cos\tau - \frac{\theta^2\epsilon}{2} + \theta^2 \epsilon
        \left[\frac{1}{2} + \cot^2\chi\right] \cos 2\tau
        \enq
(Here $P_\star = 2\pi I_3/L$ and $\varpi_p = I_3\omega_p /L = \epsilon
\Lambda_3$.) Note that a harmonic arises from geometrical effects.

\subsection{Precession Around the Principal Axis Corresponding to the Smallest Moment
of Inertia ($k^2 > 1$)}

We will now look at this case in more detail. The solution given through Eq.
(\ref{landau}) is still valid. However, it will be mathematically and
computationally convenient to carry out a transformation of the Jacobian elliptic
functions with $k^2 > 1$ into functions with parameter $1/k^2 < 1$. (That, in the
limit $\Lambda_1 = 1$ and $\Lambda_3 = 0$, we get $\omega_p = 0$ whence $\tau =
0$ and $k^2 = \infty$, provides further motivation.) Then, we can write the
general solution as,
        \beq
        &&L_1 = - \Lambda_1 \dn \hat\tau \nonumber \\
        &&L_2 = - \Lambda_3 \sqrt{1 + \hat{e}^2} \sn \hat\tau \\
        &&L_3 = \Lambda_3 \cn \hat\tau \nonumber
        \enq
where, $\hat\tau = \tau k$, $\hat{e}^2 = 1/e^2$ and the parameter of the elliptic
functions is now $\hat k^2 = 1/k^2$. We can also define a new frequency from Eq.
(\ref{tau}),
        \beq
        &&\hat\omega_p = \omega_p k = \sqrt{\frac{(I_2-I_1)(2EI_3-L^2)}{I_1I_2I_3}}
        \enq
Through an appropriate redefinition of axes, the solution can be expressed in a
form identical to the $k^2 < 1$ case, except that now $\hat{I}_1 > \hat{I}_2 >
\hat{I}_3$. Define a new right-handed coordinate basis for the rotating frame,
        \beq
        &&\m{\hat{e}}_1 = - \m{e}_3 \nonumber \\
        &&\m{\hat{e}}_2 = - \m{e}_2 \\
        &&\m{\hat{e}}_3 = - \m{e}_1 \nonumber
        \enq
Let $\hat{\Lambda}_1 = \Lambda_3$ and $\hat{\Lambda}_3 = \Lambda_1$, then the
components of the angular momentum can be expressed as,
        \beq
        &&\hat{L}_1 = -L_3 = - \hat{\Lambda}_1 \cn \hat\tau \nonumber \\
        &&\hat{L}_2 = -L_2 = \hat{\Lambda}_1 \sqrt{1 + \hat{e}^2} \sn \hat\tau \\
        &&\hat{L}_3 = -L_1 = \hat{\Lambda}_3 \dn \hat\tau \nonumber
        \enq
The precession is now clockwise, as can also be verified from Euler's equation.
$\hat{\Lambda}_i$ have exactly the same form as before, in terms of the new
moments of inertia,
        \beq
        &&\hat{\Lambda}_1 = \sqrt{\frac{\hat{I}_1 (L^2 - 2E\hat{I}_3)}
        {L^2(\hat{I}_1 - \hat{I}_3)}} \nonumber \\
        &&\hat{\Lambda}_2 = \sqrt{\frac{\hat{I}_2 (L^2 - 2E\hat{I}_3)}
        {L^2(\hat{I}_2 - \hat{I}_3)}} \\
        &&\hat{\Lambda}_3 = \sqrt{\frac{\hat{I}_3 (2E\hat{I}_1 - L^2)}
        {L^2(\hat{I}_1 - \hat{I}_3)}} \nonumber
        \enq
So do $\hat{\omega}_p$, $\hat{e}^2$ and $\hat{k}^2$, as can be verified from the
equations above. We have thus transformed the problem from a ``$k^2 > 1$ case for
an $I_3 > I_1$ body'' into a ``$\hat{k}^2 < 1$ case for an $\hat{I}_1 >
\hat{I}_3$ body'', which should not be surprising.

The equations for the Euler angles (Eqs. (\ref{euler})) remain of the same form,
with the exception of $\cos\hat\psi$. This is effectively a sign change, $\tau
\rightarrow -\hat{\tau}$, in the argument,
        \beq
        &&\hat{\zeta}(\hat\tau) = \zeta(-\hat\tau)
        \hspace{0.6cm} \mbox{ whence } \hspace{0.6cm}
        \frac{d\hat\zeta}{d\hat\tau} = - \frac{d\zeta(-\hat\tau)}{d\hat\tau}
        \enq

One must be careful with Eq. (\ref{intphi}) as well, where there is also a sign
change due to the fact that now $\hat{I}_1 > \hat{I}_3$,
        \beq
        &&\frac{L(\hat{I}_3-\hat{I}_1)}{\hat{\omega}_p \hat{I}_1\hat{I}_3} =
        - \frac{\sqrt{1+\hat{e}^2}}{\hat{\Lambda}_3} =
        - \frac{\hat{\Lambda}_2}{\hat{\Lambda}_1\hat{\Lambda}_3}
        \enq
These two effects add up to modify the function $f_n$ defined through Eq.
(\ref{func}),
        \beq
        &&\hat{f}_n (\hat\tau) = - f_n (-\hat\tau)
        \enq

\section{The Wobble and Beam Swing Angles}

We will define the wobble ($\theta$) and beam swing ($\vartheta$) angles according to
(see Fig. \ref{angles}),
        \beq
        &&\cos\theta \ = \ \hL\cdot\hat\m{z} \ = \ \Lambda_3 \dn\tau \\
        &&\cos\vartheta \ = \ \hL\cdot\hat\m{b} \ = \ - b_1\Lambda_1\cn\tau
        - b_2\Lambda_2\sn\tau + b_3\Lambda_3\dn\tau
        \label{beam}
        \enq
By definition, the wobble angle is equivalent to Euler's angle $\theta$ and is constant
for an axisymmetric star. The beam swing angle is related to the angle between the beam
and the observer. It could exceed $90^\circ$, but since the pulse will have a limited
angular size, there is a restriction on how much it can vary throughout a precession
period. Otherwise, the beam will leave the observer's line of sight. Therefore, the
span of the beam swing angle serves as a constraint. The angle can be further
restricted by imposing the conditions for an interpulse.

We define the widest span as $\Delta\vartheta = \vartheta_{max} - \vartheta_{min}$.
Then the constraint is that this be smaller than some value $\Delta\vartheta_{max}$,
which is estimated based on information about the pulse width and shape. Note that the
beam angle depends on four parameters: the two angles determining the orientation of
the pulse, and any two of $k^2$, $e^2$ and $\lambda = \Lambda_1/\Lambda_3$.

\section{Period Residuals for the Spindown Torque}

When external torques are present Euler's equation becomes,
        \beq
        &&\frac{d\m{L}}{dt} + \m{\Omega}\times\m{L} = \m{N}
        \enq
Taking the dot product with the angular momentum, we get the equation governing
the evolution of its magnitude,
        \beq
        &&\frac{d L}{dt} = \hat\m{L}\cdot\m{N}
        \label{angulaar1}
        \enq
If we now substitute $\m{L} = L\hat\m{L}$ in Euler's equation, we get, after some
rearrangement,
        \beq
        &&L\frac{d\hL}{dt} + L^2 (\m{I}^{-1} \hL) \times \hL = \m{N} -
        (\hL \cdot \m{N}) \hL
        \label{angulaar2}
        \enq
which governs the evolution of the orientation of the angular momentum. These two
are the basic equations that need to be solved. Of course, only three (of the
total of four components) are independent equations.

In the classical rotating magnetic-dipole model of pulsars, the angular momentum
is lost to radiation. The electromagnetic torque for a spherical rigid star in
vacuum is (Davis and Goldstein, 1970),
        \beq
        &&\m{N}_{vac} = - \frac{2\mu^2\Omega^3}{3c^3}
        \hat\m{b}\times(\hat\m{\Omega}\times\hat\m{b}) =
        - \frac{2\mu^2\Omega^3}{3c^3} \bra \hat\m{\Omega} -
        (\hat\m{\Omega}\cdot\hat\m{b})\hat\m{b} \ket
        \enq
Note that the torque vanishes when $\hO$ and $\hb$ are aligned. However, the
pulsar is not in a perfect vacuum, and is surrounded by a magnetosphere.
Therefore, there should be loss of angular momentum even when these two vectors
are aligned. We will therefore adopt a general spindown torque of the form,
        \beq
        &&\m{N}_{sd} = - N_o \bra\hat\m{\Omega} - a
        (\hat\m{\Omega}\cdot\hat\m{b})\hat\m{b}\ket
        \enq
where $a$ is a dimensionless parameter that measures the relative importance of
the two components. The amplitude of the spindown torque can be estimated from
observed values of the spindown time, and is small. We will be interested in a
particular example (PSR B1828--11) where the spindown time is,
        \beq
        &&t_{sd} \sim \frac{L}{N_o} \sim 10^5 \mbox{ yrs}
        \enq
Compare this with the observed precession period for the same pulsar,
        \beq
        &&P_p = \frac{2\pi}{\Omega_p} \sim \frac{I_3}{\eps L} \sim 1 \mbox{ yr}
        \enq
The ratio of the two gives,
        \beq
        &&\frac{t_{sd}}{P_p} \sim \frac{\eps L^2}{I_3 N_o} \sim 10^5
        \enq
The second term in Eq. (\ref{angulaar2}) has a magnitude of $\eps L^2 / I$,
therefore the RHS of that equation is quite negligible for the case of interest
(as will be discussed below).

The above form of the torque is true for spherical stars. This is nevertheless a good
approximation, given how small $N_o$ and $\eps$ are. In fact, we will neglect all
combinations of $N_o$ with $\eps$. This is equivalent to taking $\hO \simeq \hL$ within
all torque terms, since the angle between these two vectors is of the order of
$\epsilon$. We therefore have,
        \beq
        &&\m{N} \ = \ - N_o \bra\hL - a(\hL\cdot\hb)\hb\ket \\
        &&\frac{d L}{dt} \ = \ \hL\cdot\m{N} \ = \
        - N_o \bra 1 - a (\hL\cdot\hb)^2 \ket \label{angutte} \\
        &&L\frac{d\hL}{dt} + L^2 (\m{I}^{-1} \hL) \times \hL \ = \
        \m{N} - (\hL\cdot\m{N}) \hL
        \ = \ N_o a (\hL\cdot\hb) \bra \hb - (\hL\cdot\hb) \hL \ket
        \enq
Loss of energy (or angular momentum, given through the second equation) now
clearly requires that $a \leq 1$. Finally, we will also neglect any time
dependence within $N_o$ itself.

The third equation demands careful thought. In component form,
        \beq
        &&L \frac{d}{dt} \mateen L_1 \\ L_2 \\ L_3 \matend +
        \frac{\eps L^2}{I_1} \mateen L_2 L_3 (1-s) \\ - L_1 L_3 \\ L_1 L_2 s
        \matend = N_o a \cos\vartheta \mateen b_1 - L_1 \cos\vartheta \\
        b_2 - L_2 \cos\vartheta \\ b_3 - L_3 \cos\vartheta \matend
        \enq
where $L_i$ are the components of $\hL$, $s = (I_2 - I_1)/(I_3 - I_1)$, $\cos\vartheta =
\hL\cdot\hb$, and we have already neglected second order terms in $\eps$. As long as the
star is sufficiently non-spherical and the angular momentum is sufficiently misaligned
with the body $z$ axis, we can neglect the RHS, as it causes changes in the orientation
of the angular momentum smaller (by many orders of magnitude) than the second term. In
other words, $\eps$ and $\Lambda_1$ are small but not zero. (Keep in mind that there is
no precession if either one is zero.) Also, $s$ cannot be too close to unity, i.e. $e^2$
cannot be exceptionally large.

The same cannot be done in the RHS of the equation for the magnitude of the
angular momentum, as it is the only term we have. Incidentally, setting $N_o = 0$
would take us back to the torque-free precession case.

In order to write the equations in a dimensionless form, let's divide all sides
by a frequency $\omega_p$, defined in accordance with Eq. (\ref{tau}),
        \beq
        &&\omega_p (t) = \frac{\eps L(t) \Lambda_3}{I_3\sqrt{1+e^2}}
        \enq
but where the magnitude of the angular momentum is no longer constant. Also
define,
        \beq
        &&d\tau = \omega_p(t) dt
        \label{salve1}
        \enq
which, for a constant $\omega_p$, reduces to the familiar form of the torque-free
case. Now, the differential equations become, after some rearrangement,
        \beq
        &&\frac{d L}{d\tau} \ = \ \hL\cdot\m{N} / \omega_p \ = \
        - (N_o / \omega_p) \bra 1 - a (\hL\cdot\hb)^2 \ket \label{salve2} \\
        &&\frac{d\hL}{d\tau} + (L/\omega_p) (\m{I}^{-1}\hL)\times\hL \ = \ 0
        \enq
Since $L/\omega_p$ is time-independent, the second equation has exactly the same
solution as before, except that $\tau$ is now different, and given through a
differential equation on its own. In other words, $L_i$'s remain of the same
form. Thus, we only need to solve Eqs. (\ref{salve1}) and (\ref{salve2}).

Define a new dimensionless constant,
        \beq
        &&\varpi_p = I_3\omega_p/L =
        \eps \Lambda_3 / \sqrt{1+e^2} \la \eps
        \enq
and let's write,
        \beq
        &&L \ = \ L_o [ 1 - \ell (\tau) ] \nonumber \\
        &&\m{N} \ = \ - N_o \m{n}(\tau) \\
        &&t \ = \ [ \tau + \delta(\tau) ] / \omega_{po} \nonumber
        \enq
where $\omega_{po} = \varpi_p L_o/I_3$. The differential equations now become,
        \beq
        &&\frac{d\ell}{d\tau} = \left( \frac{I_3 N_o}{\varpi_p L_o^2} \right)
        \frac{\hL\cdot\m{n}}{1-\ell} \hspace{0.8cm} \mbox{ and } \hspace{0.8cm}
        \frac{d\delta}{d\tau} = \frac{\ell}{1-\ell}
        \label{diffel1}
        \enq
It's worth noting that we make no assumptions in these substitutions.

If we finally define one more dimensionless constant,
        \beq
        &&\tilde\Gamma_{sd} = \frac{I_3 N_o}{\varpi_p L_o^2}
        \enq
and let $\ell = \tilde\Gamma_{sd} \tilde\ell$ and $\delta = \tilde\Gamma_{sd}
\tilde\delta$, then the two equations can be written as,
        \beq
        &&\frac{d\tilde\ell}{d\tau} =
        \frac{\hL\cdot\m{n}}{1-\tilde\Gamma_{sd}\tilde\ell}
        \hspace{0.8cm} \mbox{ and } \hspace{0.8cm}
        \frac{d\tilde\delta}{d\tau} =
        \frac{\tilde\ell}{1-\tilde\Gamma_{sd}\tilde\ell}
        \label{diffel2}
        \enq
For the pulsar that we discuss here, we have,
        \beq
        &&\tilde\Gamma_{sd} \sim \frac{I_3 N_o}{\eps L_o^2} \sim \frac{P_p}{t_{sd}}
        \sim 10^{-5}
        \enq
This means that one may safely ignore the denominators of the two equations, thus
further simplifying the results,
        \beq
        &&\frac{d\tilde\ell}{d\tau} = \hL\cdot\m{n}
        \hspace{0.8cm} \mbox{ and } \hspace{0.8cm}
        \frac{d\tilde\delta}{d\tau} = \tilde\ell \label{salved}
        \enq
Note that $\hL\cdot\m{n} = 1-a(\hL\cdot\hb)^2 \geq 0$, thus assuring that $L$ is
monotonically decreasing (as required by loss of angular momentum).

\subsection{Time of Arrival Residuals}

Since $L_i$ remain of the same form, with the only difference being that $\tau$
is now determined through a differential equation (Eq. (\ref{salve1})), the Euler
angles remain the same (Eqs. (\ref{euler})). However, due to the time dependence
of $L$, it is more convenient to express $\phi$ as a function of $\tau$, and we
need to replace Eq. (\ref{intphi}) by,
        \beq
        &&\phi(\tau) \ = \ \phi_o + \int_0^\tau \frac{d\tau}{\varpi_p}
        \left( 1 + \frac{\eps}{1+e^2\sn^2\tau} \right)
        \ = \ \phi_o + \frac{\tau}{\varpi_p} + \frac{\eps}{\varpi_p}\int_0^\tau
        \frac{d\tau}{1+e^2\sn^2\tau}
        \enq
Thus, all we have to do is to replace $Lt_n/I_3$ by $\tau_n/\varpi_p$ on the LHS
of Eq. (\ref{timen}). The period (given through Eq. (\ref{period})) remains the
same as well, as does the calculation of $d\tau_n/dn$. In fact, we run into
trouble only with the period residuals, since the magnitude of the angular
momentum is now changing. Define,
        \beq
        &&\Delta P_{res} = P_n - \frac{2\pi I_3}{L_o} \hspace{0.8cm} \mbox{ and }
        \hspace{0.8cm} \Delta P_n = P_n - \frac{2\pi I_3}{L}
        \enq
$\Delta P_n$ is formally the same as the torque-free case. However, observations
give us only information about $\Delta P_{res}$. In practice, one first
determines the period $(P_\star)$ at some epoch $(t_o)$, and then finds the
period derivative ($\dot{P}_\star$) which is the secular term attributed to
spindown, and subtracts both contributions, so that the residuals are then given
through,
        \beq
        &&\Delta P_{res} = P(t) - P_\star - \dot{P}_\star \ \! (t - t_o)
        \enq
Consider the difference between the two definitions above,
        \beq
        &&\Delta P_{res} - \Delta P_n = 2 \pi I_3 \left(
        \frac{1}{L} - \frac{1}{L_o} \right) = \frac{2\pi I_3 \tilde\Gamma_{sd}}{L_o}
        \frac{d\tilde\delta}{d\tau}
        \enq
where we make use of Eqs. (\ref{diffel1}) and (\ref{diffel2}). We thus get,
        \beq
        &&\frac{\Delta P_{res}}{P_\star} = \frac{\varpi_p f_n}{1 - \varpi_p f_n} +
        \frac{\tilde\Gamma_{sd}\tilde\ell}{1 - \tilde\Gamma_{sd}\tilde\ell}
        \approx \varpi_p f_n + \tilde\Gamma_{sd} \tilde\ell
        \label{timensd}
        \enq
where $P_\star = 2\pi/\Omega_\star = 2\pi I_3/L_o$. The first term is the
geometric effect ($\Delta P_{ge}/P_\star \approx \varpi_p f_n$) and the second
term is the spindown term ($\Delta P_{sd}/P_\star \approx
\tilde\Gamma_{sd}\tilde\ell$). There are still secular terms present in the
spindown term that need to be subtracted. This will be taken care of below. The
relative amplitude of these two terms cannot be simply determined, and it is
possible that either one is dominant, or that they are comparable.

We now turn our attention to the calculation of $\tilde \ell$. From Eq.
(\ref{salved}) we have,
        \beq
        &&\tilde \ell = \int_0^\tau \hL\cdot\m{n} \ d\tau
        \hspace{0.6cm} \mbox{ where, } \hspace{0.6cm}
        \hL\cdot\m{n} = 1 - a (\hL\cdot\hb)^2
        \enq
and,
        \beq
        &&\hL\cdot\hb = - b_1\Lambda_1\cn\tau - b_2\Lambda_2\sn\tau
        + b_3\Lambda_3\dn\tau = \cos\vartheta
        \enq
Carrying out the integrals of the Jacobian elliptic functions, and substituting
the values of $k^2$ and $\Lambda_2$, we get, after some rearrangement,
        \beq
        &&\int_0^\tau (\hL\cdot\hb)^2 \ d\tau
        \ = \ \frac{\Lambda_3^2}{e^2} \left[ b_2^2 (1 + e^2)
        - b_1^2 k_1^2 \left] \tau
        + \frac{\Lambda_3^2}{e^2} \right[ b_1^2 - b_2^2(1+e^2)
        + b_3^2 e^2 \right] E(\mbox{am }\tau,k) \label{ldotb} \\
        && \hspace{2.24cm} + \ \frac{2b_1b_2\Lambda_3^2\sqrt{1+e^2}}{e^2}(1 - \dn\tau)
        - 2\Lambda_1\Lambda_3 \left[ b_2b_3\sqrt{1+e^2} (1 - \cn\tau)
        + b_1b_3\sn\tau \right] \nonumber
        \enq
where we have introduced the complementary parameter $k_1^2 = 1 - k^2$.
$E(\mbox{am}\tau,k)$ is the Legendre elliptic integral of the second kind, and
am$\tau$ is the Jacobi amplitude, am$\tau = \sin^{-1}\sn\tau$. Note that we have
implicitly assumed that at the zero of time, the angular momentum is in the $xz$
plane of the body frame. This is a non-trivial assumption, and in general one
does not have the freedom of randomly setting the initial orientation of the
angular momentum. Therefore, in general we would have $\hL= \hL(\tau - \tau_o)$,
where $\tau_o$ is the phase offset and is an additional parameter, and the
overall result would be to replace $\tilde\ell(\tau)$ above by $\tilde\ell(\tau -
\tau_o) - \tilde\ell(-\tau_o)$, where $\tilde\ell(-\tau_o)$ is just a constant.
For simplicity, we will continue to assume $\tau_o = 0$ in the rest of our
derivations, but the general case should be kept in mind.

The oscillatory part of $\tilde\ell$, after secular terms have been removed, is
given through,
        \beq
        &&\Delta\tilde\ell = \tilde\ell - \langle \hL\cdot\m{n} \rangle \tau
        \enq
The average is carried over a precession period, $2\tilde\pi = 4F(\pi/2,k)$,
        \beq
        &&\langle \hL\cdot\m{n} \rangle = 1 - a \langle (\hL\cdot\hb)^2 \rangle
        \label{ldotn}
        \enq
where, using Eq. (\ref{ldotb}) we get,
        \beq
        &&\langle (\hL\cdot\hb)^2 \rangle \ = \ \frac{1}{2\tilde\pi} \int_0^{2\tilde\pi}
        (\hL\cdot\hb)^2 \ d\tau
        \ = \ \frac{\Lambda_3^2}{e^2} \left[ b_2^2 (1 + e^2)
        - b_1^2 k_1^2 \left]
        + \frac{\Lambda_3^2}{e^2} \right[ b_1^2 - b_2^2(1+e^2)
        + b_3^2 e^2 \right] \frac{E(\pi/2,k)}{F(\pi/2,k)} \label{ldotbav}
        \enq
and we have made use of the relations $\mbox{am}(2\tilde\pi) = 2\pi$ and
$E(2\pi,k) = 4 E(\pi/2,k)$. We thus get,
        \beq
        &&\Delta\tilde\ell
        = a\langle(\hL\cdot\hb)^2\rangle \tau - a \int_0^\tau (\hL\cdot\hb)^2 \ d\tau
        \enq
which can now be used in Eq. (\ref{timensd}) to calculate the time of arrival
residuals,
        \beq
        &&\frac{\Delta P_{sd}}{P_\star} \approx \tilde\Gamma_{sd}
        \Delta\tilde\ell
        \label{dpsd}
        \enq
We will find it convenient to express this equation in the following form,
        \beq
        &&\Delta\tilde\ell / a = c_1 (1 - \cn\tau)
        + c_2 \sn\tau + \frac{c_3}{k^2} (1 - \dn\tau) +
        \frac{c_4}{k^2} \bra \frac{E(\pi/2,k)}{F(\pi/2,k)}
        \ \tau - E(\mbox{am }\tau,k) \ket
        \label{spindown}
        \enq
where,
        \beq
        &&c_1 = 2\Lambda_2\Lambda_3 b_2b_3 \mbox{ , } \hspace{0.3cm}
        c_2 = 2\Lambda_1\Lambda_3 b_1b_3 \mbox{ , } \hspace{0.3cm}
        c_3 = - 2\Lambda_1\Lambda_2b_1b_2 \hspace{0.3cm} \mbox{ and } \hspace{0.3cm}
        c_4 = \Lambda_1^2 [ b_1^2 - b_2^2(1+e^2) + b_3^2 e^2 ] \nonumber
        \enq
These coefficients are related to each other through,
        \beq
        &&c_4 = -\frac{c_2c_3}{2c_1} + \frac{c_1c_3}{2c_2} - \frac{c_1c_2}{2c_3}k^2
        \label{cik}
        \enq

It is also interesting to note that $\Delta\tilde\ell$ has a non-zero average
over a precession period. The residuals may have non-zero average depending on
when and how the period and its derivatives are calculated. This becomes
particularly important when calculating the time of arrival residuals, which can
be obtained by integrating the period residuals, and if the period residuals have
a constant term, then the time of arrival residuals will have a linear term.
Therefore, in calculating the time of arrival residuals one will have to subtract
any constant terms from the period residuals.

\subsection{Amplitude of the Residuals}

Consider the period derivative, which is given through the secular terms in
$\tilde \ell$,
        \beq
        &&\dot P_\star  = \eta P_\star\tilde\Gamma_{sd}\omega_{p}
        \label{pdot}
        \enq
where $\eta = \langle \hL\cdot\m{n} \rangle = 1 - a c_o$ and $c_o = \langle
\cos^2\vartheta \rangle$. The amplitude of the period residuals thus becomes,
from Eq. (\ref{dpsd}),
        \beq
        &&A = aP_\star \tilde\Gamma_{sd} =
        \frac{a\dot P_\star}{\omega_p (1 - ac_o)} = \frac{aA_o}{1 - ac_o}
        \hspace{0.6cm} \mbox{where} \hspace{0.6cm}
        A_o = \frac{P_\star}{2\omega_p t_{sd}}
        \label{amplitude}
        \enq
and $t_{sd} = P_\star/2\dot P_\star$ is the spindown time. Recall that $a$
measures the strength of the oscillating part of the spindown torque, and must be
$\leq 1$.

For PSR B1828--11 the period is 405.04 ms, the precession period is about 511
days, and the spindown time is 0.11 Myr, so that we get $A_o \simeq 409.95$ ns.

\subsection{The Axisymmetric Body}

For an axisymmetric star $e^2 = k^2 = 0$, but $\lambda = k/e \neq 0$. Due to the
symmetry we can set $b_2 = 0$ by shifting the zero of time through some phase
$\tau_o$. (Note that the same cannot be done in a triaxial body, where we chose
to fix the axes according to the principal moments of inertia.) In this case Eq.
(\ref{spindown}) reduces to the form,
        \beq
        &&\Delta\tilde\ell / a = \frac{\lambda}{1+\lambda^2} \left[ \sin 2\chi
        \sin (\tau - \tau_o) -
        \frac{\lambda}{4} \sin^2\chi \sin 2(\tau - \tau_o) \right]
        \enq
Note that the non-linearity of the dipole contribution of the torque naturally
brings in a harmonic. The period residuals are then,
        \beq
        &&\frac{\Delta P_{sd}}{P_o} = \tilde\Gamma_{sd} \Delta\tilde\ell
        \hspace{0.6cm} \mbox{ where } \hspace{0.6cm} \tilde\Gamma_{sd} =
        \frac{N_o}{\omega_p L_o} = \frac{1}{2 \omega_p \tau_c}
        \enq
Here $\tau_c$ is the characteristic time,
        \beq
        &&\tau_c = \frac{3c^3I_3}{4\mu^2\Omega_o^2}
        \enq
It turns out that, for the axisymmetric case, the geometric term is quite
negligible compared to the spindown term, for the range of physical parameters of
interest (Jones \& Andersson, 2001; Link \& Epstein, 2001).

To convert our result for period residuals ($\Delta P/P_o$) into residuals of the
derivative of the angular velocity ($\Delta\dot\Omega/\Omega_o$) given by Link \&
Epstein, we make use of,
        \beq
        &&\Delta\dot P = \frac{d\Delta P}{dt} = \omega_p \frac{d\Delta P}{d\tau}
        \hspace{0.6cm} \mbox{ and } \hspace{0.6cm}
        \frac{\Delta\dot\Omega}{\Omega_o} = - \frac{\Delta\dot P}{P_o}
        \enq
which indeed gives the correct results, together with the initial phase
difference of $\pi$ between the two definitions,
        \beq
        &&\frac{\Delta\dot\Omega}{\Omega_o} =
        \frac{a \lambda}{2\tau_c(1+\lambda^2)} \left[ - \sin 2\chi
        \cos (\tau - \tau_o) + \frac{\lambda}{2} \sin^2\chi \cos
        2(\tau - \tau_o) \right]
        \enq

There is one difference between the two derivations and that is the presence of
the coefficient $a$ which measures the strength of the spindown torque. With the
addition of this new element, the number of unknowns increases to three ($a$,
$\lambda$ and $\chi$), while a fit to data will yield only two coefficients
($a_1$ and $a_2$; $\tau_o$ does not contain any further information). This
implies that there is a certain level of freedom in the choice of the physical
coefficients. Let's denote the fitting function by $f$,
        \beq
        &&f = a_1 \sin (\tau - \tau_o) - a_2 \sin 2(\tau - \tau_o)
        \enq
Then, the relations between the coefficients of this function and the physical
parameters that we actually seek would be,
        \beq
        &&a_1 = \frac{a \lambda \sin 2\chi}{1+\lambda^2} \hspace{0.6cm} \mbox{ and }
        \hspace{0.6cm} a_2 = \frac{a \lambda^2 \sin^2\chi}{4(1 + \lambda^2)}
        \enq
Define $\lambda = \tan\theta$, and the ratio of the two coefficients gives,
        \beq
        &&\tan\chi\tan\theta = \frac{8a_2}{a_1} \label{tantan}
        \enq
It is also possible to express $\chi$ and $\theta$ as functions of $a$. However,
as it turns out, the range of the physical parameters is severely restricted by
the beam swing angle constraint, which in the axisymmetric case is given through,
        \beq
        &&\Delta\vartheta = 2\min\left(\chi,\theta\right) < \Delta\vartheta_{max}
        \enq
This forces one of the two angles to be small (which will have $\tan_1 < 0.09$ even if
we let $\Delta\vartheta_{max} = 10^\circ$); while Eq. (\ref{tantan}) ensures that the
other remains very close to $90^\circ$. (The ratio of the coefficients is found to be
$a_2/a_1 \sim 0.4$ for the data used by Link \& Epstein (2001). This yields the
condition $\tan_2 > 36$, i.e. the second angle has to be larger than $88^\circ$, in
accordance with previous findings.)

\section{Statistical Inference}

Denote the set of parameters by $\vec x$. Then the pdf for the parameters can be
calculated as, by Bayes's theorem,
        \beq
        &&P(\vec x|D,M)={P(\vec x|M)P(D|\vec x,M)\over P(D|M)}
        \label{post}
        \enq
where $D$ stands for data, $M$ stands for the model and also takes into account
any other information that is available on the problem apart from data (in this
case the beam swing angle, which we impose as a restriction on the parameter
space). $P(\vec x|M)$ is the prior probability for the parameters; $P(D|\vec
x,M)$ is the likelihood; and $P(D|M)$ is effectively a normalization constant.

To find the pdf for a certain parameter, or a subset of parameters, we integrate Eq.
(\ref{post}) over the remaining parameters. For that we need to know the likelihood. For
each data point $y_i$ at time $t_i$, we have a theoretical prediction $f_i = f(t_i|\vec
x,M)$. For well-known uncertainties with Gaussian distribution, we would then have,
        \beq
        &&P(D|\vec x,M) = \prod_i(\sigma_i\sqrt{2\pi})^{-1}
        \exp\left[-{(y_i - f_i)^2\over 2\sigma_i^2} \right]
        \enq
If we assume that the error bars are not well-determined and rescale them through
some number $F$, the above equation becomes,
        \beq
        &&P(D|\vec x,M)=\prod_i(F\sigma_i\sqrt{2\pi})^{-1}
        \exp\left[-{(y_i - f_i)^2\over 2F^2\sigma_i^2}\right]
        \enq
and we regard $F$ as an additional parameter. We have to introduce a prior for
$F$. Since we do not want it to depend much on the endpoints, we take it to be
flat over $d\ln F$, i.e. proportional to $dF/F$, and integrate over all values of
$F$. Define,
        \beq
        &&\sum_i{(y_i - f_i)^2\over 2\sigma_i^2}\equiv \chi^2_o(\vec x)
        \enq
whence we get, for $d$ data points,
        \beq
        &&\int_0^\infty{dF~P(D|\vec x,M)\over F}
        = \left( \prod_i{1\over\sigma_i\sqrt{2\pi}} \right)
        \int_0^\infty{e^{-\chi_o^2/F^2} d F\over F^{d+1}}
        \propto \bra\chi_o^2(\vec x)\ket^{-d/2}
        \enq
where we have dropped anything that does not depend on the remaining parameters
$\vec x$, including integrals that give constants, products of the original
sigmas, and factors of $\sqrt{2\pi}$. Thus, our final result is,
        \beq
        &&P(\vec x|D,M) \ \propto \ P(\vec x|M) \bra\chi_o^2(\vec x)\ket^{-d/2}
        \enq
where the first term is the prior probability, and the constant of
proportionality can be computed from the condition that the final pdf is
normalized to one.

\label{lastpage}

\end{document}